\documentclass{PoS}

\PoS{PoS(LAT2005)021}

\usepackage{epsfig}

\title{Theoretical issues with\\ staggered fermion simulations}

\ShortTitle{Theoretical issues with staggered fermion simulations}

\author{\speaker{Stephan D\"urr}\thanks{Based on work done in collaboration
        with Christian Hoelbling (Wuppertal) and Urs Wenger (DESY Zeuthen).}\\
        Bern University, ITP\\
        Sidlerstrasse 5, 3012 Bern, Switzerland\\
        E-mail: \email{durrATitp.unibe.ch}}

\abstract{The legality of the ``rooting trick'' in dynamical staggered fermion
simulations is discussed, i.e.\ whether the theory with the Boltzmann weight
$\det^{1/4}(D_\mathrm{st})$ yields the right continuum limit. Since the problem
is unsolved, pieces of evidence in favor and against are collected and
examined.}

\FullConference{XXIIIrd International Symposium on Lattice Field Theory\\
                25-30 July 2005\\
                Trinity College, Dublin, Ireland}

\newcommand{\al}{\alpha}
\newcommand{\be}{\beta}
\newcommand{\ga}{\gamma}
\newcommand{\de}{\delta}

\newcommand{\ze}{\zeta}
\newcommand{\et}{\eta}
\renewcommand{\th}{\theta}

\newcommand{\la}{\lambda}

\newcommand{\si}{\sigma}
\newcommand{\ta}{\tau}

\newcommand{\ch}{\chi}
\newcommand{\ps}{\psi}
\newcommand{\om}{\omega}

\newcommand{\Mpi}{M_{\pi}}
\newcommand{\Fpi}{F_{\pi}}

\newcommand{\pa}{\partial}
\newcommand{\ovr}{\over}
\newcommand{\til}{\tilde}

\renewcommand{\dag}{^\dagger}
\newcommand{\<}{\langle}
\renewcommand{\>}{\rangle}
\newcommand{\gaf}{\gamma_5}
\newcommand{\nab}{\nabla\!}
\newcommand{\lap}{\triangle}
\newcommand{\trc}{\mr{tr}}
\newcommand{\psb}{\bar{\psi}}
\newcommand{\psd}{\psi^\dagger}
\newcommand{\chb}{\bar{\chi}}
\newcommand{\zed}{\zeta^\dagger}

\newcommand{\bdm}{\begin{displaymath}}
\newcommand{\edm}{\end{displaymath}}
\newcommand{\bea}{\begin{eqnarray}}
\newcommand{\eea}{\end{eqnarray}}
\newcommand{\beq}{\begin{equation}}
\newcommand{\eeq}{\end{equation}}

\newcommand{\mr}{\mathrm}

\newcommand{\ri}{\mathrm{i}}
\newcommand{\cd}{\!\cdot\!}
\newcommand{\Nf}{{N_{\!f}}}%{N_\mr{f\,}}
\newcommand{\Nt}{{N_{t}}}%{N_\mr{t\,}}
\newcommand{\MeV}{\,\mr{MeV}}
\newcommand{\GeV}{\,\mr{GeV}}

\newcommand{\fm}{\,\mr{fm}}
\newcommand{\const}{\mr{const}}

\newcommand{\Dst}{D_\mr{st}}
\newcommand{\Dov}{D_\mr{ov}}
\newcommand{\Dca}{D_\mr{ca}}
\newcommand{\Dstm}{D_{\mr{st},m}}
\newcommand{\Dovm}{D_{\mr{ov},m}}
\newcommand{\Dcam}{D_{\mr{ca},m}}

\begin{document}

%%%%%%%%%%%%%%%%%%%%%%%%%%%%%%%%%%%%%%%%%%%%%%%%%%%%%%%%%%%%%%%%%%%%%%%%%%%%%%%

\section{Introduction}

Is ``staggered QCD'' really QCD, or is it just a model of QCD ?
This is the question addressed in this note, and it is worth emphasizing two
points right now: the issue is whether the staggered approach yields the right
continuum limit and the question is asked at the non-perturbative level.

The staggered action is not doubler-free; the Kogut-Susskind construction
manages to reduce the fermion multiplicity from 16 to 4 \cite{Kogut:1974ag}.
To simulate QCD with $\Nf\!=\!2\!+\!1$ dynamical quarks staggered studies
employ the ``rooting trick'', i.e.\ the Boltzmann weight used in the simulation
is
\beq
e^{-S_\mr{eff}}=
\det\nolimits^{1/2}(D_{\mr{st},m_{ud}})\,
\det\nolimits^{1/4}(D_{\mr{st},m_s})\,
e^{-S_G}
\label{seff_root}
\eeq
where $m_{ud}\!=\!(m_u\!+\!m_d)/2$ is an average up and down quark mass,
$m_s$ is kept at the physical strange quark mass and $S_G$ denotes the gauge
action.
Throughout, I use the abbreviation $\Dstm\!=\!\Dst\!+\!m$ for the massive
staggered Dirac operator.
The ``rooting trick'' is the way how the unphysical ``tastes'' (see below)
are excised from the sea sector of the theory, and the issue of debate is
whether the effective action may be re-expressed in terms of an artificial
undoubled formulation which is \emph{local}.
Ideally, this construction would hold exactly on typical backgrounds, but in
view of the first point emphasized above it might be OK if one can prove that
a slight mismatch in the determinants is just a cut-off effect.
I will explain below why it is so useful to have a local underlying theory.
On the level of the effective action the question is whether multi-plaquette
correlators have unphysical cuts in the limit where the lattice spacing $a$
is sent to zero.

The staggered action not being doubler-free has an impact on the valence
sector, too.
Here, sophisticated reduction techniques have been developed (already in the
quenched era) to project out the desired spinor$\!\otimes\!$taste combination
from a given staggered operator.
The problem is that the way how the unphysical tastes are excised from the
valence sector is (in general) slightly different from the rooting procedure
employed in the sea sector.
Evidently, this means that at finite lattice spacing the theory is
\emph{not unitary\/}.
Again, this does not necessarily lead to a disaster, if one can show that
the non-unitarity is a genuine cut-off effect.

Personally, I tend to see the locality as the crucial issue and the unitarity
as a point of lesser severity.
This is so, because there are reasons to believe that a solution to the first
problem would automatically provide a bypass to the second one, as sketched
below.

\bigskip

The controversy, as I see it, is thus a debate whether the rooting procedure in
the sea sector would introduce spurious degrees of freedom which corrupt, at
the non-perturbative level, the long-range properties of the effective action
in such a way that something survives the continuum limit.
As of now, the answer is not known.
All I can provide is an attempt to explain basic facts needed to understand the
origin of the problem and then collect, in a second step, pieces of evidence in
favor and against the legality of the staggered approach.
It turns out that the evidence ``in favor'' outnumbers the one ``against'' by
far, but in the end the sheer number has no say.

A key issue in practical applications of staggered fermions is to offset the
taste symmetry breaking as much as possible.
In the course of the simulation this is done by using improved/filtered
varieties of the staggered action which may reduce the taste splitting quite
drastically (see below).
In the analysis of pseudo-scalar meson properties Staggered Chiral Perturbation
Theory (SXPT) is used, an effective field theory that allows one to
parameterize and thus ``take out'' the leading taste breaking effects.
Many of the pieces of evidence to be presented below will use one or the other
strategy to tame the taste symmetry breaking.
Still, from a fundamental viewpoint it is clear that either one of these
techniques is immaterial -- if the staggered approach is correct, it will yield
the right continuum limit for arbitrary observables without any of these.

Obviously, there are two ways how the controversy could play out.
One possible outcome is realized, if someone manages to give an unambiguous
(mathematical) proof that QCD with $\Nf\!=\!2\!+\!1$ staggered quarks is in the
right universality class.
The other option is, of course, that someone manages to find a single
observable where he/she can demonstrate that the staggered answer, after a
continuum extrapolation, is wrong.

\bigskip

Before getting into details, let me comment on one trap to be avoided.
It is true that some of the more recent studies with $\Nf\!=\!2\!+\!1$
dynamical staggered quarks agree (in particular in observables relating to
$\Upsilon$ spectroscopy) with experiment to a precision that is currently not
matched by another fermion discretization (see e.g.\ \cite{Davies:2003ik}).
Still, from a conceptual viewpoint this is no help in deciding whether the
staggered approach is fundamentally correct or ``epicycle physics''.
In particular the agreement of a mixed QCD/electroweak observable with
experiment is not really an argument, since what we want to find out is whether
there is a contribution from beyond-the-SM physics in such observables.
What comparing to experiment is supposed to tell us is whether the Lagrangian
relevant to a problem is the one we stick into our code, and for this the
method itself has to be from first principles.
Hence, whether QCD with $\Nf\!=\!2\!+\!1$ staggered fields is fundamentally
correct is an issue that needs to be solved theoretically -- by pure thought
or with the help of CPU power.

\bigskip

The plan is to first review some elements of the staggered action and taste
representation.
The question about the legality of the ``rooting trick'' is then exposed
as a locality problem that would be solved if one could find a
``candidate'' operator that satisfies certain criteria.
In the free theory the search has ended successfully,
and I discuss the four constructions known to date.
For the interacting theory I assess pieces of evidence that come
from studying spectral and eigenmode properties of the staggered action in 4D.
For many observables in 2D it is easy to reach good statistical precision and
this leads to a somewhat more mixed picture in the Schwinger model.
Still, just comparing the staggered and the overlap contribution to the
effective action [i.e.\ correlating $-{1\ovr2}\log\det(\Dstm)$ and
$-\log\det\Dovm$ in a scatter plot for $m\!>\!0$] lets one ask whether $\Dovm$
might actually represent (one version of) the desired ``candidate'' action.
Finally, a few words on what can be learned from analyzing data with
Staggered Chiral Perturbation Theory are in place.
The problem being unsolved, I refrain from drawing any conclusion and end
with a summary.

%%%%%%%%%%%%%%%%%%%%%%%%%%%%%%%%%%%%%%%%%%%%%%%%%%%%%%%%%%%%%%%%%%%%%%%%%%%%%%%

\section{Review: Staggered action and taste representation}

The staggered action derives from the naive fermion discretization
\beq
S_\mr{na}={a^3\ovr2}\sum_{x,\mu}\psb(x)\,\ga_\mu\,
[U_\mu(x)\ps(x\!+\!\hat\mu)-U_\mu\dag(x\!-\!\hat\mu)\ps(x\!-\!\hat\mu)]
\eeq
which is known to lead to 4 (in 2D) or 16 (in 4D) fermions in the continuum
($2^d$ species in $d$ dimensions).
The observation by Kogut and Susskind was that the $x$-dependent field
transformation  $\ps(x)=\ga(x)\ch(x),\psb(x)=\chb(x)\ga(x)\dag,
\ga(x)=\ga_1^{n_1}\ga_2^{n_2}\ga_3^{n_3}\ga_4^{n_4}$ with $x\!=\!na$ and the
staggered phase factor $\et_\mu(x)=(-1)^{\sum_{\mu<\nu}n_\nu}$ brings it to the
form
\beq
S_\mr{na/st}={a^3\ovr2}\sum_{x,\mu}\chb(x)\,\et_\mu(x)\,
[U_\mu(x)\ch(x\!+\!\hat\mu)-U_\mu\dag(x\!-\!\hat\mu)\ch(x\!-\!\hat\mu)]
\label{def_stag}
\eeq
where the 2 (in 2D) or 4 (in 4D) components of $\ch$ (in general: $2^{d/2}$)
manifestly decouple.
Therefore, by just ``downgrading'' $\ch$ to be a 1-component field, the action
(\ref{def_stag}) is changed to the staggered one which yields only
$\Nt=2^{d/2}$ fermions in the continuum.
The Kogut Susskind procedure thus manages to ``thin out'' degrees of
freedom, albeit at the price of distributing/intertwining ``spinor'' and
``taste'' (the latter being the modern word to address the full fermion content
of a staggered field).
An important point is that the $SU(\Nt)_A$ taste symmetry in the formal
continuum limit reduces to the abelian invariance
$\ch(x)\to e^{\ri\th_{\!A\,}\et_5(x)}\ch(x),
\chb(x)\to e^{-\ri\th_{\!A\,}\et_5(x)}\chb(x)$ of the massless action at finite
lattice spacing, and this is good enough to prevent additive mass
renormalization.
However, further exact ``thinning'' by purely algebraic means is impossible,
since the Kogut Susskind procedure has exhausted the eigenvalue degeneracy of
the naive action (the latter has only $2^{d/2}$-fold degenerate eigenvalues on
typical backgrounds, not $2^d$-fold, as one might have naively guessed).

\bigskip

Modern staggered simulations use an individual staggered field for each
continuum flavor to be studied, i.e.\ QCD with $\Nf\!=\!2\!+\!1$ quarks
requires the $\ch$-fields $u,d,s$.
The taste structure of physical operators is non-trivial
\cite{Patel:1992vu,Sharpe:1994dc}.
The underlying taste identification may be attempted in momentum space
\cite{Sharatchandra:1981si} or in coordinate space
\cite{Gliozzi:1982ib,Duncan:1982xe,Kluberg-Stern:1983dg}.
The latter version builds on the hypercubic decomposition
$\ch(x,x\!+\!a\hat1,...,x\!+\!a\hat1\!+\!a\hat2\!+\!a\hat3\!+\!a\hat4) \to q(X)$
which collects all spinor/taste information in a $2^4$ sublattice (of the
original one) into a ``masternode'', i.e.\ a point in the blocked lattice with
spacing $b\!=\!2a$.
Denoting these coarse points with uppercase symbols, the free action reads
\beq
S_\mr{st}=b^4\sum_{X,\mu}\bar q(X)\;
[\nab_\mu(\ga_\mu\!\otimes\!I)-{b\ovr2}\lap_\mu(\gaf\!\otimes\!\ta_\mu\ta_5)]
\;q(X)
\label{stag_hypdec}
\eeq
where $q(X)$ is a 16-component ``quark field'' (in 4D) and
the first and second derivative operate on the blocked lattice, i.e.\
$(\nab_\mu q)(X)={q(X+b\hat\mu)-q(X-b\hat\mu)\ovr 2b}$ and
$(\lap_\mu q)(X)={q(X+b\hat\mu)-2q(X)+q(X-b\hat\mu)\ovr b^2}$.
The tensor structure in (\ref{stag_hypdec}) refers to spinor$\otimes$taste,
where $\ta_\mu\!=\!\ga_\mu^*,\!\ta_5\!=\!\ga_5$, i.e.\ another set of gamma
matrices is used to map out the taste algebra.
The main observation is that in this basis the troublesome taste changing
interactions stem from a dimension 5 Wilson-type term which is ``skewed'' in
spinor$\otimes$taste space (and hence not positive semidefinite).

The case of interest is the interacting theory, and here the question is
whether all taste changing effects would go away, in the continuum limit,
without leaving any trace.
Of course, from a formal viewpoint this seems almost obvious, since it is a
dimension 5 piece
\footnote{As worked out by Luo the $O(a)$ term in (\ref{stag_hypdec}) can be
shifted to $O(a^2)$ by a change of basis \cite{Luo:1996vt}.}
in (\ref{stag_hypdec}) that mediates the taste changes, and irrelevant
operators have no anomalous dimensions.
As pointed out in \cite{DeGrand:2003xu} the whole issue is a peculiar
exchange-of-limits question.
If the formal continuum limit is justified (and the view that naturally derives
from the taste basis thus adequate), then everything is fine, indeed.
The problem is just that one would need to prove first that the staggered
action yields the correct continuum limit to be sure that adopting this
viewpoint is not misleading.

One way to think about taste changing interactions is to notice that in the
interacting theory individual taste (or spinor) components of a $q$-field
still sit out there in the little $2^4$ hypercube attached to a blocked
``masternode''.
They see slightly different local gauge fields (therefore
spinor$\otimes$taste projection operators must be gauged, e.g.\ the staggered
$\gaf$ is a 4-link operator).
Equivalently, in the momentum space view each spinor/taste component has a
reduced Brillouin zone with extension $2\pi/b\!=\!\pi/a$ in each direction
and this means that hard gluons with one component $\sim\!\pi/a$ will flip one
taste to another one.
In any case it is clear that the identification of physical and staggered
flavor may only work, if taste interactions are minimized (ideally:
eliminated).

\begin{figure}
\epsfig{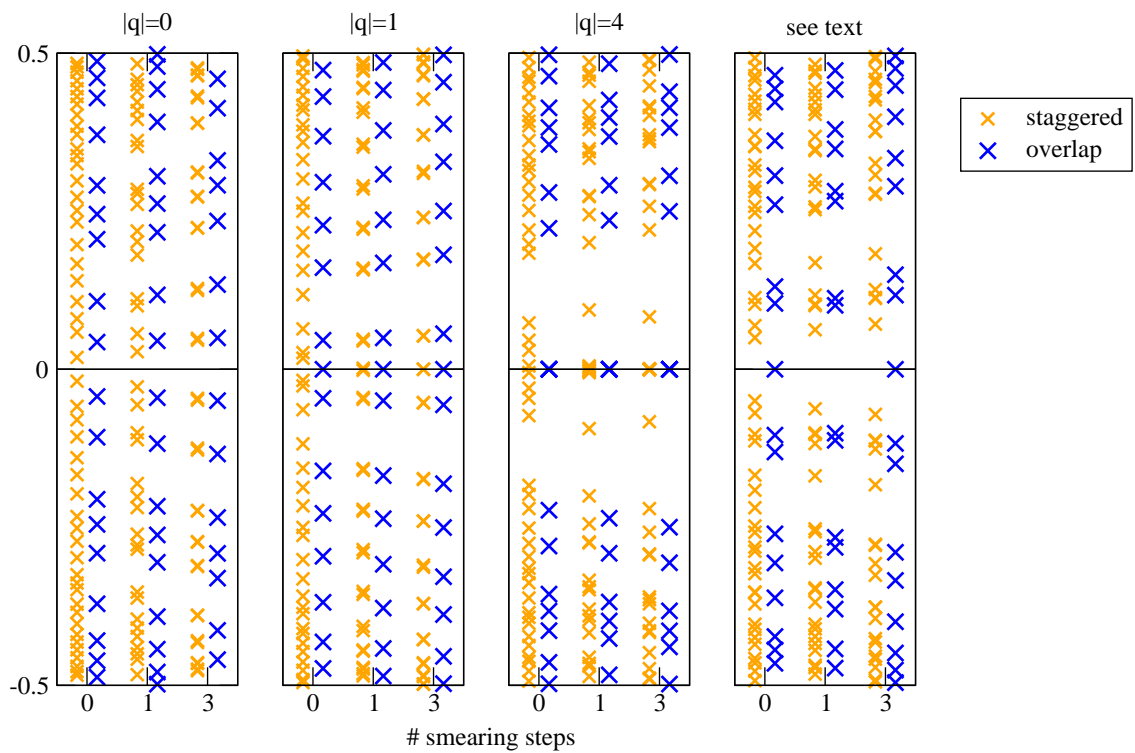}
\vspace{-2mm}
\caption{Staggered and overlap eigenvalues on 4 selected configurations in 2D.
On benevolent configurations the improved taste symmetry achieved through
filtering results in a 2-fold (4-fold in 4D) near-degeneracy and the right
number of near-zero modes. To aid comparison the low-energy overlap spectrum
has been mapped onto the imaginary axis with a stereographic projection.
Figure from \cite{Durr:2003xs}.}
\label{fig_spec_sm}
\end{figure}

Everything below will hinge on this taste symmetry (which at finite lattice
spacing is only approximate), and Fig.\,\ref{fig_spec_sm} is an attempt to
convey a down-to-earth view.
If the taste symmetry were exact the staggered eigenvalue spectrum would show
a 2-fold (in 2D) or 4-fold (in 4D) degeneracy.
On typical configurations smearing (i.e.\ using a filtered fermion operator,
see below) is a convenient tool (read: cheaper than decreasing $a$) to achieve
a rather good $\Nt$-fold near-degeneracy, and this is the basis of the
rooting procedure.
In other words, if the symmetry/degeneracy were exact, it would be conceivable
that one might reduce the degrees of freedom by another factor of $2^{d/2}$
(analogous to Kogut and Susskind, but with different technical means).
However, at finite lattice spacing there are always some configurations
where the taste symmetry is not approximately restored (rightmost panel),
and the issue concerning the ``rooting trick'' is basically the question
whether these ``bad guys'' would retain non-zero measure in the continuum.

%%%%%%%%%%%%%%%%%%%%%%%%%%%%%%%%%%%%%%%%%%%%%%%%%%%%%%%%%%%%%%%%%%%%%%%%%%%%%%%

\section{Problem: rooting versus locality}

The earliest warning regarding the legality of the ``rooting trick'' that I am
aware of came in a paper by Marinari, Parisi and Rebbi in 1981:
``On the lattice the action $S_G\!-\!{1\ovr4}\trc\log(\Dstm)$ will produce a
violation of fundamental axioms, but we expect the violation to disappear in
the continuum limit and then recover the theory with a single fermion''
\cite{Marinari:1981qf}.

As mentioned in the introduction, the main problem is the potential loss of
\emph{locality\/} at finite lattice spacing.
With any undoubled Dirac operator, e.g.\ $D\!=\!D_\mr{W}$, one has
\beq
Z|_\Nf=\int\!\!\!D[U,\!\psb,\!\ps]\;e^{-S_G[U]-\sum\!\int\!\psb D\ps}=
\int\!\!\!D[U]\,\det\nolimits^\Nf(D)\;e^{-S_G[U]}
\eeq
where the first sum extends over $\Nf$ flavors, i.e.\ an integer power of
the determinant emerges rather naturally with $\Nf$ degenerate fermions.
The question is whether one may take fractional powers if a formulation
with doublers yields too many fermions in the continuum, specifically whether
\beq
Z|_\Nf=
\int\!\!\!D[U]\;\det\nolimits^{\Nf/4}(\Dstm)\;e^{-S_G[U]}
\label{Z_nf_root}
\eeq
is a legal implementation if $\Nf$ is not a multiple of 4.
The point is that it is not clear whether the resulting effective action
$S_\mr{eff}=S_G\!-\!{\Nf\ovr4}\trc\log(\Dstm)$ receives contributions from
unphysical degrees of freedom which persist in the continuum.
In principle, such a disaster might happen with any non-zero prefactor, and
locality is thus a issue in the staggered approach for any $\Nf\!>\!0$.

The reason why locality is so important is \emph{universality\/}.
We are all used to the notion that we do not need to worry about the details
of the action used, since the continuum limit is a $2^\mr{nd}$ order phase
transition in the course of which all physical correlation lengths diverge
and any knowledge about irrelevant pieces in the action is washed out.
However, the standard theorems ensuring these nice properties build on the
fundamental action being local (at finite lattice spacing) \cite{Book_MM}.
%Furthermore, the locality or causality of arbitrary Green's function is
%also a standard prerequisite to discuss the renormalizability of a field
%theory \cite{...}.

The soundness of the effective action would be guaranteed if one could
``integrate in'' a fermion in (\ref{Z_nf_root}), i.e.\ if one could find a
legal doubler-free ``candidate'' action $\Dcam$ with \cite{Jansen:2003nt}
\beq
\det\nolimits^{1/4}(\Dstm)=
\int\!\!\!D[\psb,\!\ps]\;e^{-\int\!\psb\Dcam\ps}
\;.
\label{def_Dca}
\eeq
The attribute ``legal'' means that it satisfies the requirements
\bea
&&\til\Dca=\ri\ga_\mu p_\mu+O(p^2)
\label{req_1}
\\
&&||\Dca(x,y)||<C\,e^{-\nu|x\!-\!y|/a}\;\;
\mbox{with}\;C,\!\nu\;\mbox{independent of}\;U
\label{req_2}
\\
&&||\de\!\Dca(x,y,U)/\de U(z)||<C\,e^{-\nu\min(|x\!-\!z|,|y\!-\!z|)/a}\;\;
\;(\mbox{ditto})
\label{req_3}
\eea
where (\ref{req_2},\ref{req_3}) are designed to ensure locality in the space of
couplings and in the gauge field, and ``doubler-free'' is formalized by
demanding that $\til\Dca(p)$ is invertible at all non-zero momenta
(mod $2\pi/a$).
Of course, one might allow for cut-off effects in (\ref{def_Dca}), i.e.\ one
might relax it to
\beq
\det(\Dcam) \stackrel{a\downarrow0}{\longrightarrow}
\const\cd\det\nolimits^{1/4}(\Dstm)
\eeq
but the problem will be to prove that the difference is really just a cut-off
effect.

\bigskip

What might be a promising ``candidate'' action ?
Sometimes it is easier and more instructive to discuss first what is not
a useful strategy.
The staggered ``rooting problem'' is of this type, since it is evident that
simply taking the fourth root of $\Dstm$ is not a viable option.
Such an operator would have the right determinant, but it would violate other
requirements, in particular $\Dca\!=\!\Dst^{1/4}$ with $\Dst$ the massless
staggered operator would fail to obey (\ref{req_1}).

\begin{figure}
\begin{center}
\epsfig{file=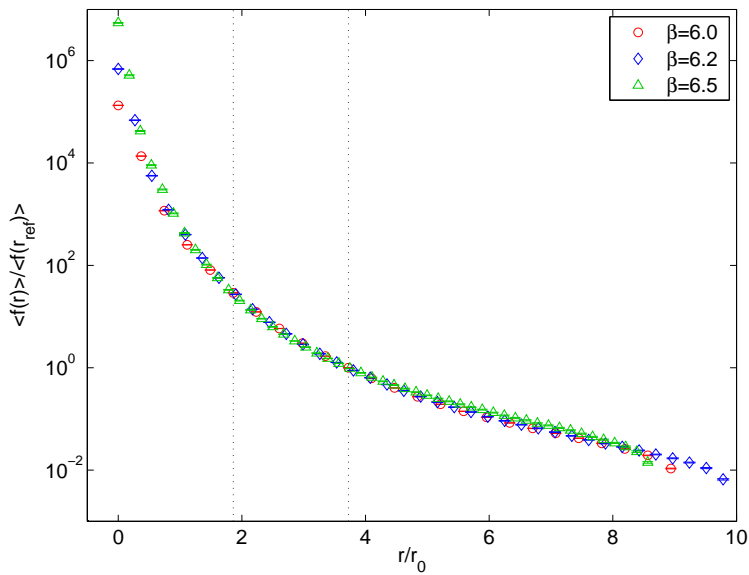,width=10cm}
\end{center}
\vspace{-8mm}
\caption{Localization function $f(r)$ in physical units versus $r/r_0$ for the
``candidate'' action of \cite{Bunk:2004br}. The local logarithmic derivative
(the effective $r_\mr{loc}$) scales and the operator is thus non-local. Figure
from \cite{Bunk:2004br}.}
\label{fig_bunk}
\end{figure}

That the technique \cite{Hernandez:1998et} is available to decide whether a
potential/proposed ``candidate'' action obeys the locality bound (\ref{req_2})
has bee shown in \cite{Bunk:2004br}.
These authors consider a slightly different ``candidate'' operator,
$\Dcam=(\Dstm\dag\Dstm^{})_\mr{ee}^{1/2}=(-\Dst^2\!+\!m^2)_\mr{ee}^{1/2}>0$,
and measure the localization function
$f(r)=\mr{sup}\{\;||\ps(x)||_2\,\big|\,||x\!-\!y||_1=r\;\}$ where
$\ps(x)=\sum\!D(x,z)\et(z)$ with $\et$ a normalized random vector at $y$.
For a local $D$ one has $f(r)\!\propto\!e^{-r/r_\mr{loc}}$, ideally with
$r_\mr{loc}\!\propto\!a$, but any $a$-dependence with
$\xi_\mr{phys}/r_\mr{loc}\!\to\!\infty$ is fine.
Employing matched lattices they find that $f(r/r_0)$ scales (and there is no
convincing exponential fall-off pattern, see Fig.\,\ref{fig_bunk}), i.e.\ the
operator is non-local.
Thus, apart from not qualifying as a fermion action since its spectrum lies
on the real axis, this ``candidate'' operator may also be ruled out for more
sophisticated reasons.
Using an improved/filtered staggered action in the above-mentioned construct
does not alter the conclusion \cite{Hart:2004sz}, which is hardly surprising,
since in the previous work the continuum limit has already been taken.

Still, the question remains: is there \emph{any\/} ``candidate'' operator
satisfying all requirements ?
Note that finding a legal 1-flavor $\Dcam$ is sufficient to show that staggered
fermions yield an acceptable effective action.
The continuum limit might be OK even if this is not the case, but in such an
event a proof that the approach is safe would -- almost for sure -- require
truly novel concepts.

%%%%%%%%%%%%%%%%%%%%%%%%%%%%%%%%%%%%%%%%%%%%%%%%%%%%%%%%%%%%%%%%%%%%%%%%%%%%%%%

\section{Free case: four constructions}

Given the importance of the issue whether a ``candidate'' action exists that
satisfies all relevant criteria, it is encouraging to see that at least in
the free theory the answer is positive and the task to construct such an
operator has been completed.
Interestingly, the solution is non-unique.

In \cite{Adams:2004mf} Adams starts from the observation that the massive
staggered operator in the taste basis (\ref{stag_hypdec}) on the blocked
lattice may be used to build the combination
\beq
\Dstm\dag \Dstm^{}=
\Big[-\nab\,^2+{b^2\ovr4}\lap^2+m^2\,\Big](I_4\!\otimes\!I_4)
\label{adams_1}
\eeq
which is simultaneously diagonal in spinor$\otimes$taste.
Analogously, a free generalized Wilson operator
$\Dcam=\nab_\mu\ga_\mu+{b\ovr2}W+m$ on the blocked lattice yields
\beq
\Dcam\dag \Dcam^{}=
\Big[-\nab\,^2+\big({b\ovr2}W+m\big)^2\,\Big](I_4\!\otimes\!1)
\label{adams_2}
\eeq
without taste structure.
With $\det(\Dstm\dag)\!=\!\det(\Dstm^{})$ and
$\det(D_{\mr{W},m}\dag)\!=\!\det(D_{\mr{W},m}^{})$ it follows that $\Dcam$
would trivially have the right determinant, if the two square brackets in
(\ref{adams_1}) and (\ref{adams_2}) agree:
\beq
{b^2\ovr4}\lap^2+m^2=\Big({b\ovr2}W+m\Big)^2
\quad\Longrightarrow\quad
\det\nolimits^{1/4}(\Dstm)=\det(\Dcam)
\label{adams_3}
\;.
\eeq
Hence, a natural ``candidate'' operator in the free theory is
$\Dcam=\nab_\mu\ga_\mu+\sqrt{{b^2\ovr4}\lap^2+m^2}$ which lives
on the blocked lattice.
The less trivial part is to verify its locality.
Using a technique developed by Fujikawa and Ishibashi he shows that the
couplings fall off exponentially with
$r_\mr{loc}=\sqrt{8a/m}\stackrel{a\downarrow0}{\longrightarrow}0$
and this means that for $m\!>\!0$ the operator is local.
%but obviously for $m\!=\!0$ the construction does not work.

\begin{figure}
\begin{center}
\epsfig{file=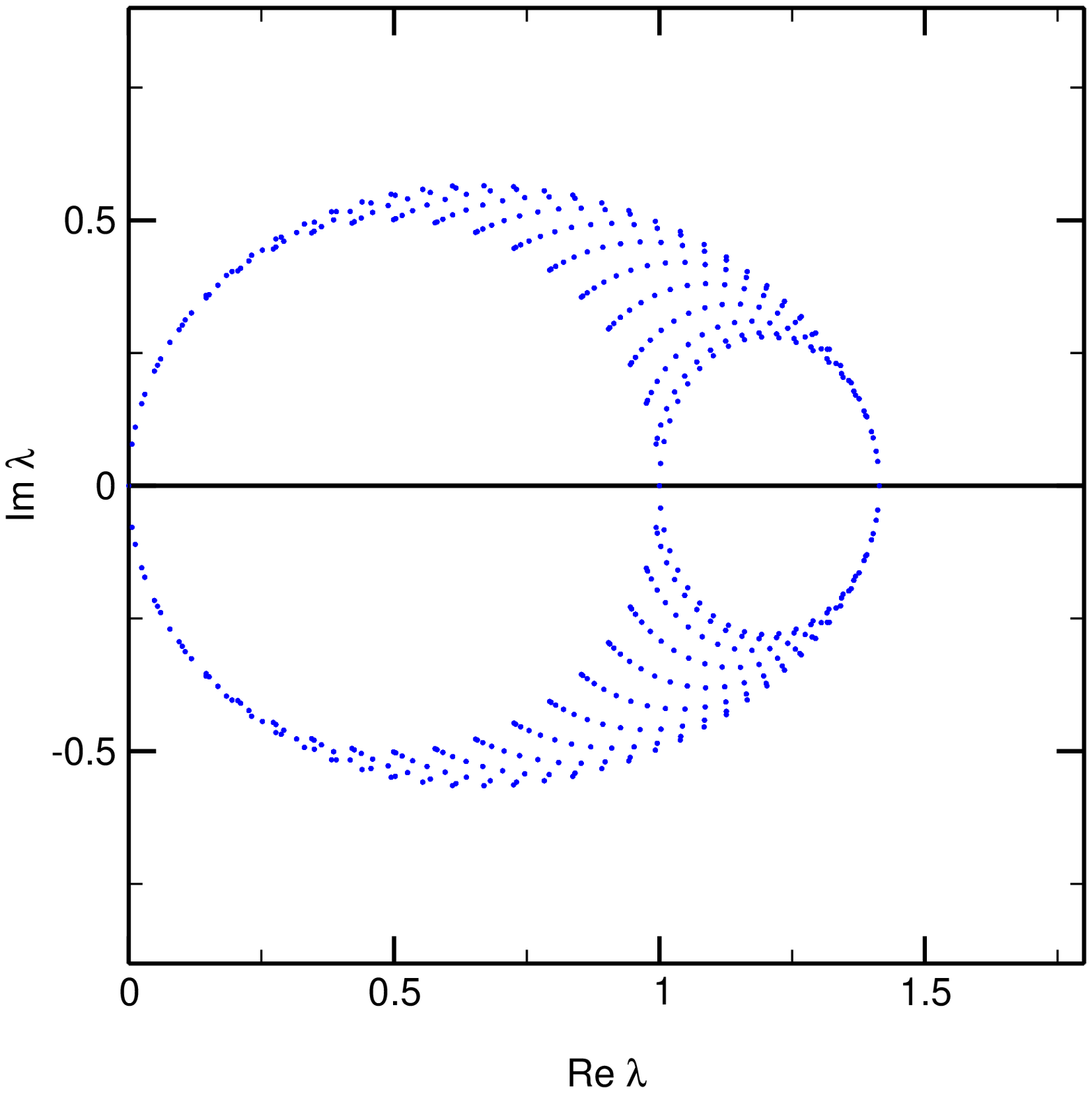,width=6cm}
\epsfig{file=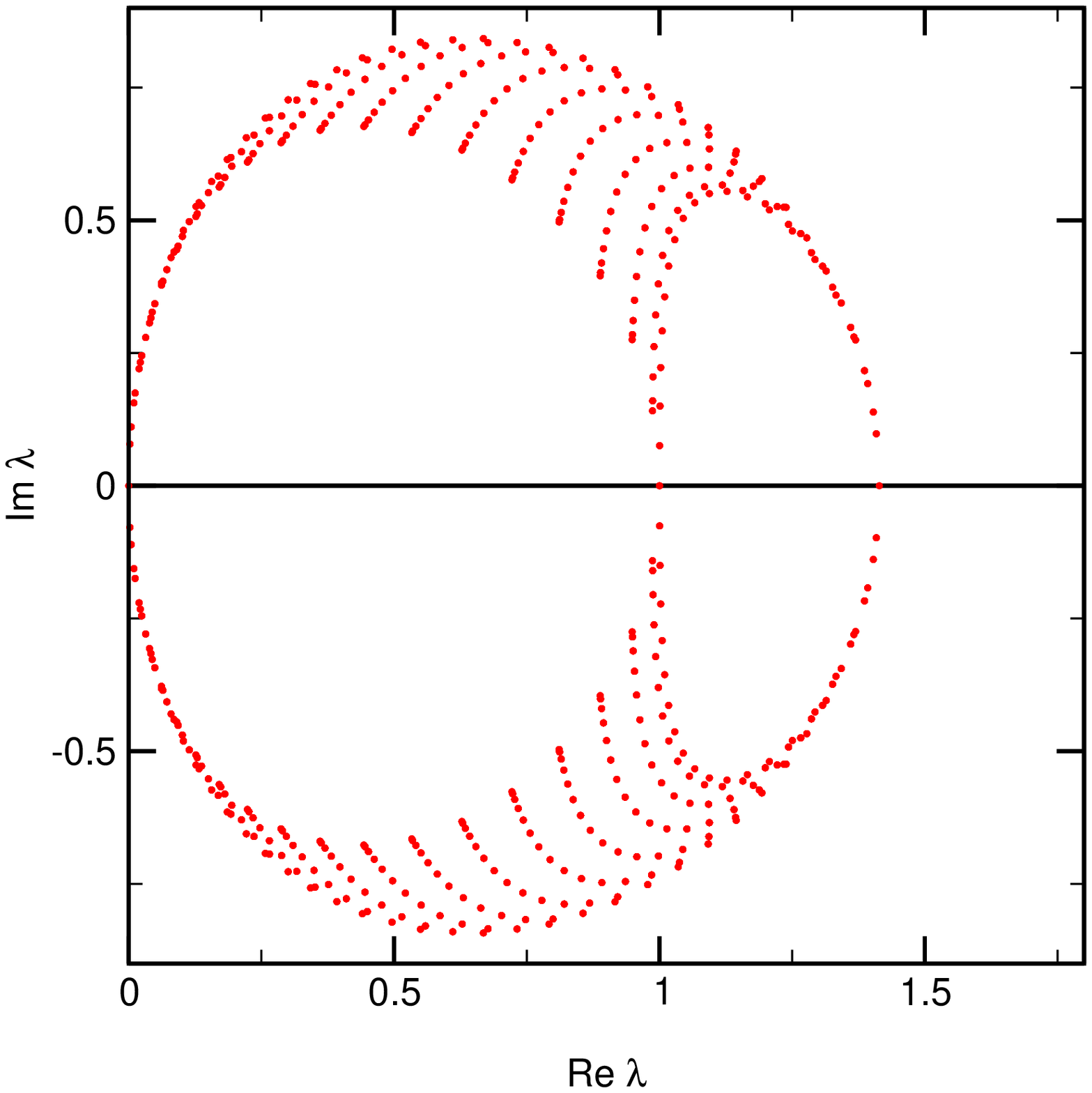,width=6cm}
\end{center}
\vspace{-8mm}
\caption{Eigenvalue spectra of two ``candidate'' operators $\Dca$ on the
blocked lattice in 2D  which reproduce $\det^{1/2}(\Dst)$ of the massless
staggered action in the free theory. On the left $\Dca$ is exclusively
optimized for locality, on the right the (standard) Ginsparg Wilson relation
is imposed as a constraint. Figure from \cite{Maresca:2004me}.}
\label{fig_marescapeardon}
\end{figure}

In \cite{Maresca:2004me} Maresca and Peardon investigate the massless case.
They start from a ``candidate'' operator $\Dca=\ri\ga_\mu P_\mu\!+\!Q$ on the
blocked lattice with local (i.e.\ not ultra-local) $P_\mu$ and $Q$.
Again, $\det(\Dca)=\det^{1/\Nt}(\Dst)$ is trivially realized if they can ensure
that $\Dca\dag \Dca^{}\!\otimes\!I_{\Nt}^\mr{taste}=\Dst\dag \Dst^{}$.
To that aim they consider the ansatz 
$P_\mu=\sum_{r\geq0}\sum_{|d|\leq r}\om_{p,\mu}^{r,d}(x,y)$
and ditto for $Q$ in terms of a tower of operators with ever larger footprint,
i.e.\ $\om_{p,\mu}^{r,d}, \om_q^{r,d}$ have range $r$.
Ensuring the main criterion does not uniquely pin down all the coefficients,
and they use this freedom to optimize $\Dca$ for locality.
This optimization is performed numerically, both in 2D ($\Nt\!=\!2$) and in 4D
($\Nt\!=\!4$).
With this strategy they make sure that $\Dca$ has the right determinant
[my understanding: up to the exact (non-topological) zero-modes], and it turns
out that the resulting operator is indeed exponentially localized.
Presumably triggered by the spectrum of their ``candidate'' operator in 2D
(see Fig.\,\ref{fig_marescapeardon}) they went on to ask whether they could
make it obey the Ginsparg Wilson relation \cite{Ginsparg:1981bj}.
Indeed, considering only $Q$ of the form $Q\!=\!-\lap\!R$ with a local $R$
makes the ``candidate'' operator satisfy
\beq
\{\Dca,\gaf\}=\Dca\,2R\gaf\,\Dca
\label{maresca_peardon_1}
\eeq
but the price to pay is that the localization range increases.

In \cite{Shamir:2004zc} Shamir uses the renormalization group (RG) framework
to deal with the free case.
The general idea is that even in the interacting theory performing $n$
blocking steps (for simplicity: $n$ coarse grainings by a factor 2)
creates a ``two cut-off'' situation where the physics accessible on the
coarse lattice is given by the final cut-off $a^{-1}/2^n$, but the taste
splitting is still given by the original $a^{-1}$.
Thus, starting from an ever finer lattice and applying more blocking steps
is a way to improve the taste symmetry (cf.\ talk by F.\,Maresca
\cite{Talk_Maresca}).
Infinitely many steps render it exact, i.e.\
$D_n\!\to\!D_\infty\!\otimes\!I_4$, while $\Dca=D_n$ after $n$ steps
satisfies $\det^{1/4}(\Dst)\!=\!\det(\Dca)\det^{1/4}(T)$
where $T$ contains only (original) cut-off excitations and should maintain the
Symanzik class, i.e.\ $\det(T)\!=\!\mr{const}\cd(1\!+\!O(a^2))$.
The massless staggered action on the original lattice satisfies
$\{D_0,(\gaf\!\otimes\!\ta_5)\}\!=\!0$.
Interestingly, the action after $n$ blocking steps satisfies a generalized
Ginsparg Wilson relation \cite{Shamir:2004zc}
\beq
\{D_n,(\gaf\!\otimes\!\ta_5)\}=D_n\,{2\ovr\al_n}(\gaf\!\otimes\!\ta_5)\,D_n
\label{shamir_1}
\eeq
where $\al_n$ is a parameter of the blocking procedure.
If (\ref{shamir_1}) were complemented with a
$(\gaf\!\otimes\!\ta_5)$-hermiticity of $D_n$, one would easily obtain
$D_n\!+\!D_n\dag\!=\!D_n{2\ovr\al_n}D_n\dag$, i.e.\ that the spectrum of
$D_n$ lies on a Ginsparg-Wilson circle.
However, it is straightforward to derive spectral properties without such
a $(\gaf\!\otimes\!\ta_5)$-hermiticity.
Start from the representation $D_n\!=\!\sum_j\la_j u_j^{} v_j\dag$ in
terms of bi-orthogonal R/L-eigenvectors, i.e.\ $v_i\dag u_k^{}\!=\!\de_{i,k}$.
Upon sandwiching the GW relation (\ref{shamir_1}) with $v_i\dag$ and $u_k^{}$,
one obtains $\la_i v_i\dag (...)u_k+v_i\dag (...)u_k \la_k=
\la_i v_i\dag {2\ovr\al_n}(...)u_k \la_k$ where $...$ stands for
$\gaf\!\otimes\!\ta_5$.
Thus, for an arbitrary pair $(i,k)$ of eigenmodes, one has
$v_i\dag (\gaf\!\otimes\!\ta_5) u_k^{}\!=\!0$
or $\la_i\!+\!\la_k\!-\!{2\ovr\al_n}\la_i\la_k\!=\!0$.
In particular for $i\!=\!k$ it follows that
$v_j\dag (\gaf\!\otimes\!\ta_5) u_j^{}\!\neq\!0$ necessarily implies
$2\la_j\!-\!{2\ovr\al_n}\la_j^2\!=\!0$.
Hence the spectrum crosses the real axis at $\la_j\!\in\!\{0,\al_n\}$.
The more general situation with a spectrum between two ``Hasenfratz-Niedermayer
circles'' (tangent to the imaginary axis) is avoided, and the reason why is
simply that Shamir uses a non-overlapping blocking kernel.

There are two more articles to be mentioned.
In \cite{Giedt:2005he} Giedt gives an exploratory discussion of the interacting
case, based on RG concepts.
In \cite{Neuberger:2004be} Neuberger uses a different setup; he recasts the
question of the validity of the ``rooting trick'' into a local field
theoretical framework in 6D.

%%%%%%%%%%%%%%%%%%%%%%%%%%%%%%%%%%%%%%%%%%%%%%%%%%%%%%%%%%%%%%%%%%%%%%%%%%%%%%%

\section{Interacting theory: $\mr{spec}(\Dst)$ in 4D}

Before discussing what can be learned from analyzing spectral properties of
the staggered Dirac operator, let me recall the concept of \emph{filtering\/}.
What is meant is simply a change of the covariant derivative in the fermion
action \cite{DeGrand:1998pr,Blum:1996uf,Lepage:1998vj,Orginos:1999cr,
Hasenfratz:2001hp,DeGrand:2002vu} through a ``fat link'' recipe like
\beq
U_\mu(x)\ps(x\!+\!\hat\mu)-\ps(x)\quad\longrightarrow\quad
U_\mu^\mr{HYP}(x)\ps(x\!+\!\hat\mu)-\ps(x)
\;.
\label{def_filter}
\eeq
The idea is that such a modification alters/diminishes cut-off effects without
changing the Symanzik class, i.e.\ a theory with $O(a^2)$ cut-off effects will
retain this characteristics, but the extrapolation slope is changed (typically
reduced).
For staggered quarks the obvious design goal of a filtering prescription is to
render $D_\mr{st}$ ``immune'' against $p\!\sim\!\pi/a$ gluons, and the question
is how efficient a given recipe is at ameliorating the taste symmetry.
For Wilson fermions the additive mass renormalization gets reduced through the
modification (\ref{def_filter}) and for related technical reasons even the
overlap operator \cite{Neuberger:1997fp} benefits from such a filtering
\cite{Durr:2004as,Durr:2005an}.

\begin{figure}
\begin{center}
\epsfig{file=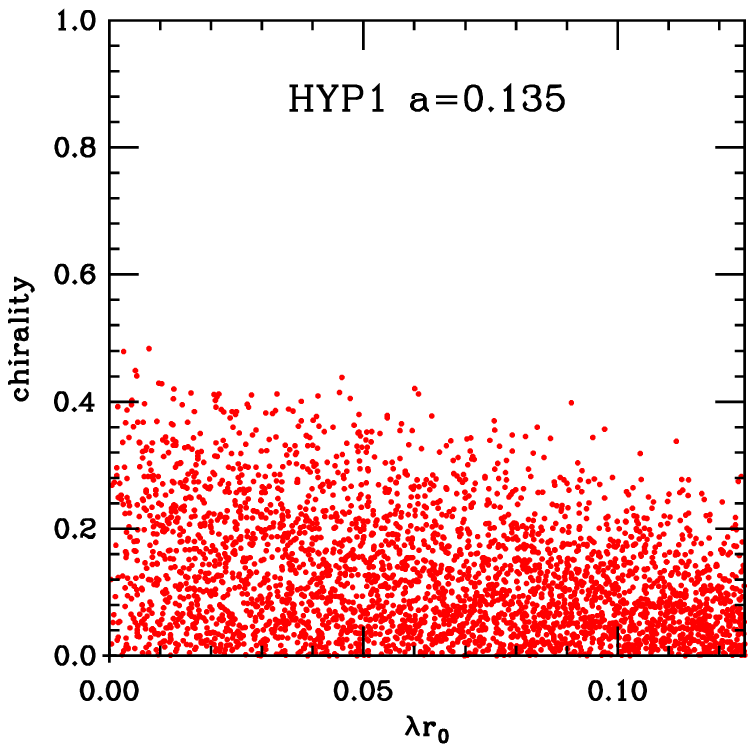,width=6cm,angle=-90}
\epsfig{file=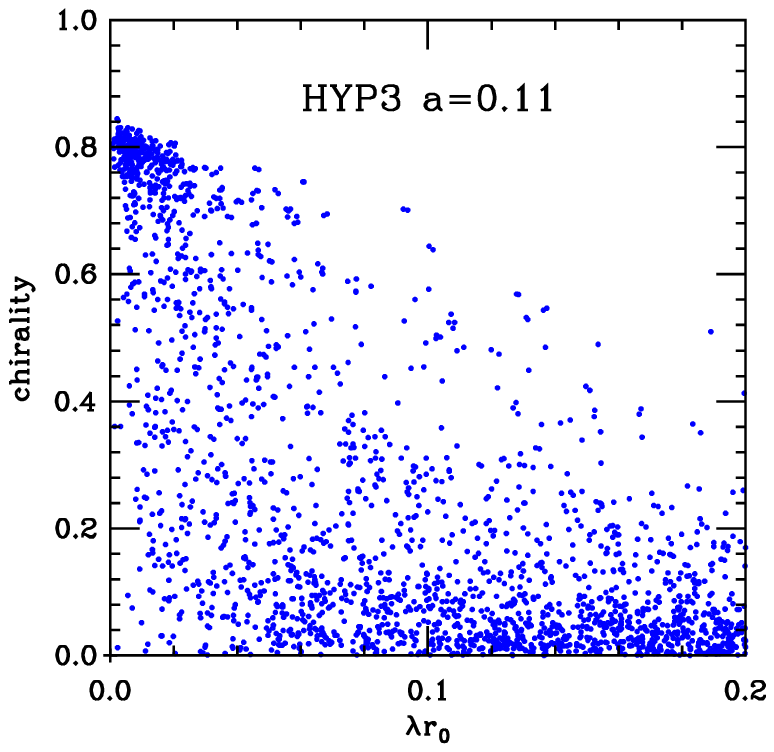,width=6cm,angle=-90}
\end{center}
\vspace{-6mm}
\caption{Chirality versus eigenvalue of staggered eigenmodes. On coarse
lattices there is no distinctive signature, but on somewhat finer lattices
rather aggressive filtering reveals continuum like features. Near-zero modes
are predominantly chiral, non-zero modes are not. Figure
from \cite{Talk_Hasenfratz}.}
\label{fig_hasenfratz}
\end{figure}

An excellent starting point is found in a talk by A.\,Hasenfratz
\cite{Talk_Hasenfratz}.
In the (formal) continuum
$D\ps\!=\!\la\ps\,\Longleftrightarrow\,D\gaf\ps\!=\!-\la\gaf\ps$, thus non-zero
modes come in complex conjugate pairs ($\la$ is purely imaginary) and have zero
chirality ($\psd\gaf\ps\!=\!0$ if $\la\!\neq\!0$), while zero modes have
non-vanishing chirality ($\zed\gaf\ze\!=\!\pm1$ if $D\ze\!=\!0$).
On coarse lattices unfiltered and modestly filtered staggered quarks show
no indication of such ``continuum like'' behavior (see left panel of
Fig.\,\ref{fig_hasenfratz}), but with $a^{-1}\!\simeq\!2\GeV$ and some
rather aggressive filtering the situation changes, on a qualitative level
at least (see right panel of Fig.\,\ref{fig_hasenfratz}).
The idea is that the same picture would emerge with the ``thin link'' action,
too, if one could push to much smaller lattice spacings (say
$a^{-1}\!\simeq\!20\GeV$ or so).
At least it becomes evident that cut-off effects may
completely mask the underlying continuum theory.

\begin{figure}
\begin{center}
\epsfig{file=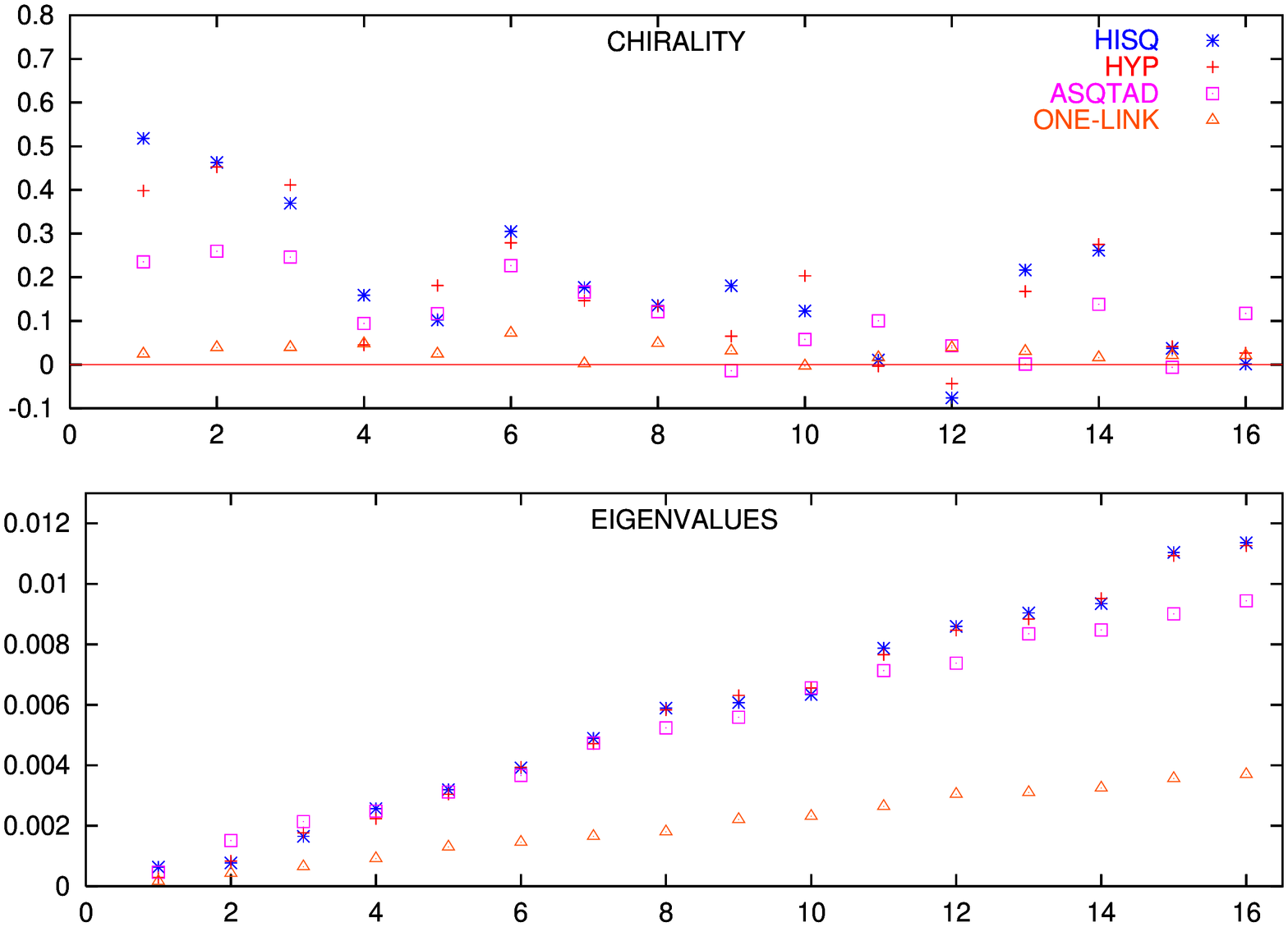,height=7.5cm,angle=-90}
\epsfig{file=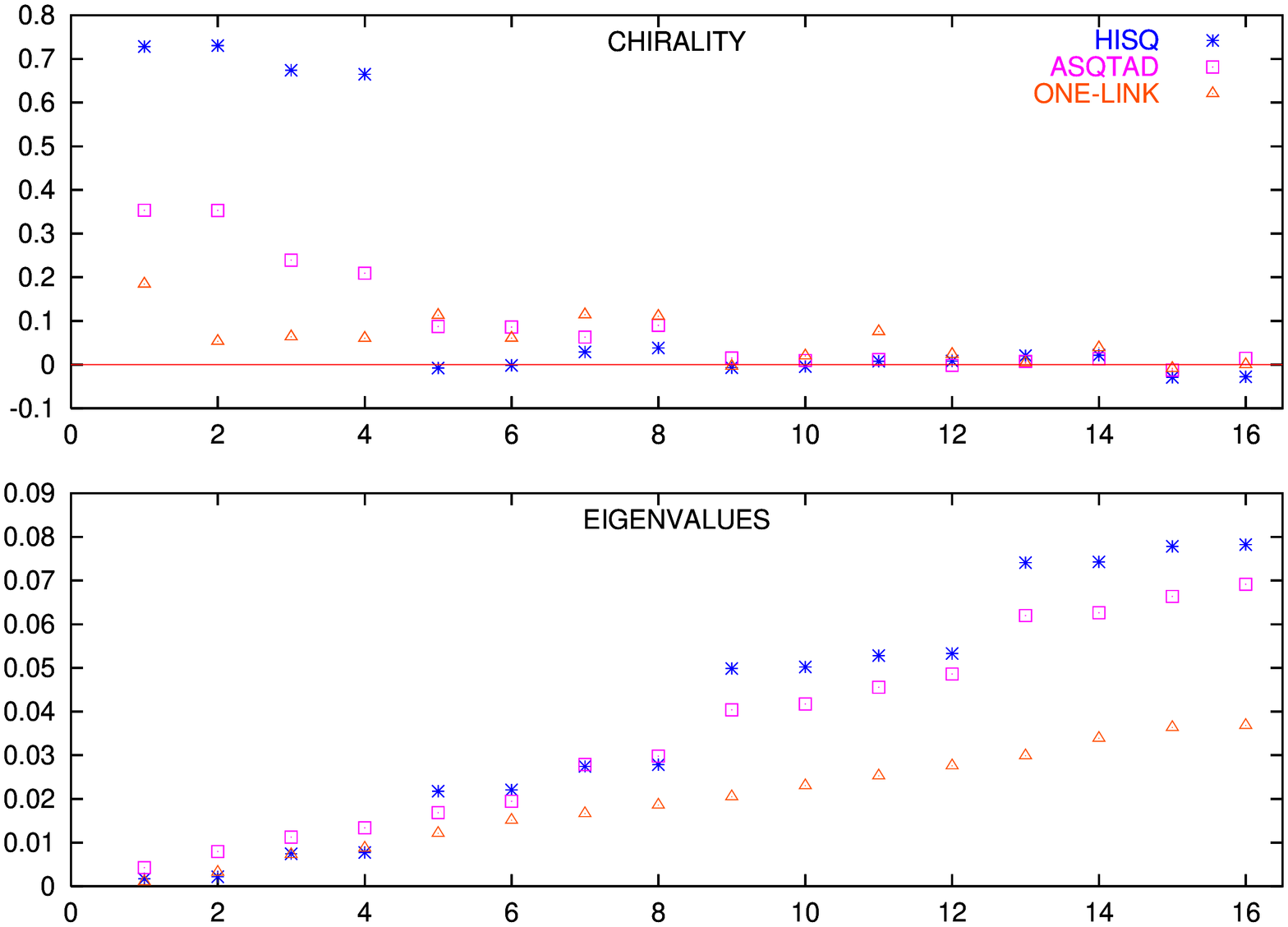,height=7.5cm,angle=-90}
\end{center}
\vspace{-4mm}
\caption{Chirality and eigenvalue versus ``serial number'' of the lowest few
staggered eigenmodes of the positive staggered half-spectrum. The left panel is
for Wilson glue, the right one for the Symanzik action, with
$a\!\simeq\!0.1\fm$ in either case. The staggered action with ``HISQ''
filtering sees $|q|\!=\!2$.
Figure from \cite{Follana:2004sz}.}
\label{fig_follana1}
\vspace{-2mm}
\begin{center}
\epsfig{file=stagissues.figs/08080808_5.79_uno.eps_mod,width=7.5cm}
\epsfig{file=stagissues.figs/16161616_6.18_uno.eps_mod,width=7.5cm}
\end{center}
\vspace{-6mm}
\caption{Same as the lower part of Fig.\,5, but with two different lattice
resolutions (keeping the physical volume fixed) and with overlap eigenvalues on
the same configuration for comparison. After rescaling with
$Z_S^\mr{st}/Z_S^\mr{ov}$ one sees a quantitative single-overlap to 
staggered-quartet agreement. Figure from \cite{Durr:2004as}.}
\label{fig_duhowe1}
\end{figure}

A more complete investigation has been carried out and presented last year by
Follana {\it et al.} \cite{Follana:2004sz,Follana:2004wp}.
The difference is that they show results for individual configurations and
compare several filtering recipes and gauge actions (at fixed lattice spacing).
Furthermore, they determine the (gluonic) topological charge $q$ of their
configurations.
With lattice spacings $a^{-1}\!\simeq\!2\GeV$ only the combination of an
improved gauge action and a highly filtered $\Dst$ allows to see a separation
into near-zero modes with non-vanishing chirality and non-zero modes without
chirality (cf.\ Fig.\,\ref{fig_follana1}).
Furthermore, the number of near-zero modes is $4|q|$ (i.e.\ $2|q|$ on either
side of the spectrum), and this means that staggered quarks satisfy an
\emph{approximate index theorem\/}.

The next step has been taken in a paper together with Christian Hoelbling and
Urs Wenger where two new features come in \cite{Durr:2004as}.
The first one is that we compare to the overlap spectrum on the same
configuration.
Apart from finding $4|q|$ staggered near-zero modes on typical backgrounds
with overlap charge $q$ we show that after rescaling with the relative
$Z_S$-factor one finds a quantitative agreement of staggered quartets with the
(non-degenerate) overlap eigenvalues (cf.\ Fig\,\ref{fig_duhowe1}).
The second new ingredient is that we compare several matched lattices (with
fixed physical volume) and investigate how the intra-quartet splitting
decreases with a finer resolution (see below).

\begin{figure}[!b]
\begin{center}
\epsfig{file=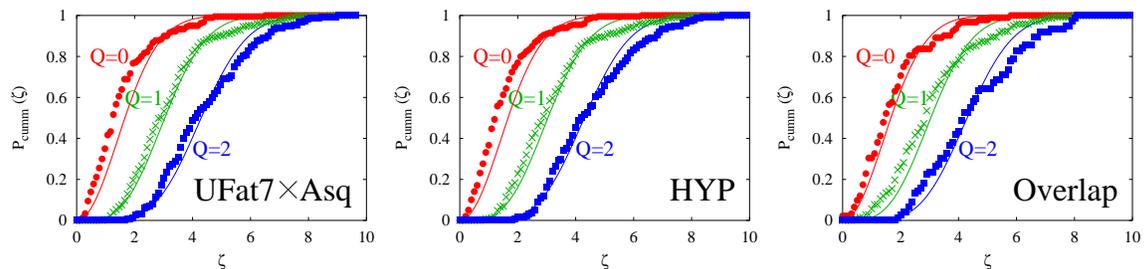,angle=-90,width=15cm}
\end{center}
\vspace{-6mm}
\caption{Cumulative eigenvalue distribution for two filtered $\Dst$ and the
``thin-link'' $\Dov$. Figure from \cite{Wong:2004nk}.}
\label{fig_wong}
\end{figure}

For quite some time it was believed that staggered quarks are insensitive to
topology and would never reproduce the sectoral random matrix theory (RMT)
predictions for the distribution/ratio of the lowest eigenvalues.
Given the correspondence to the overlap spectrum it is clear that this cannot
be true.
In \cite{Wong:2004nk} it is shown that on sufficiently fine lattices and
with filtering the agreement is just as good as with the overlap action
(see Fig.\,\ref{fig_wong}).
In \cite{Follana:2004sz} it was already shown that the ratio of low-lying
eigenvalues [i.e.\ $\<\la_i\>/\<\la_j\>$ with $\la_i$ denoting a staggered
quartet mean] satisfies the RMT prediction, up to small violations which might
be finite volume effects or remnant discretization errors.

\begin{figure}
\begin{center}
\epsfig{file=stagissues.figs/scaling.fig3.eps_mod,width=7.5cm}
\epsfig{file=stagissues.figs/scaling.fig1.eps_mod,width=7.5cm}
\end{center}
\vspace{-6mm}
\caption{Staggered near-zero eigenvalue (left) and intra-quartet splitting
(right) versus the lattice spacing squared [everything in physical units].
With filtering one might be in the scaling window.
Figure from \cite{Durr:2004as}.}
\label{fig_duhowe2}
\vspace{-2mm}
\begin{flushright}
\epsfig{file=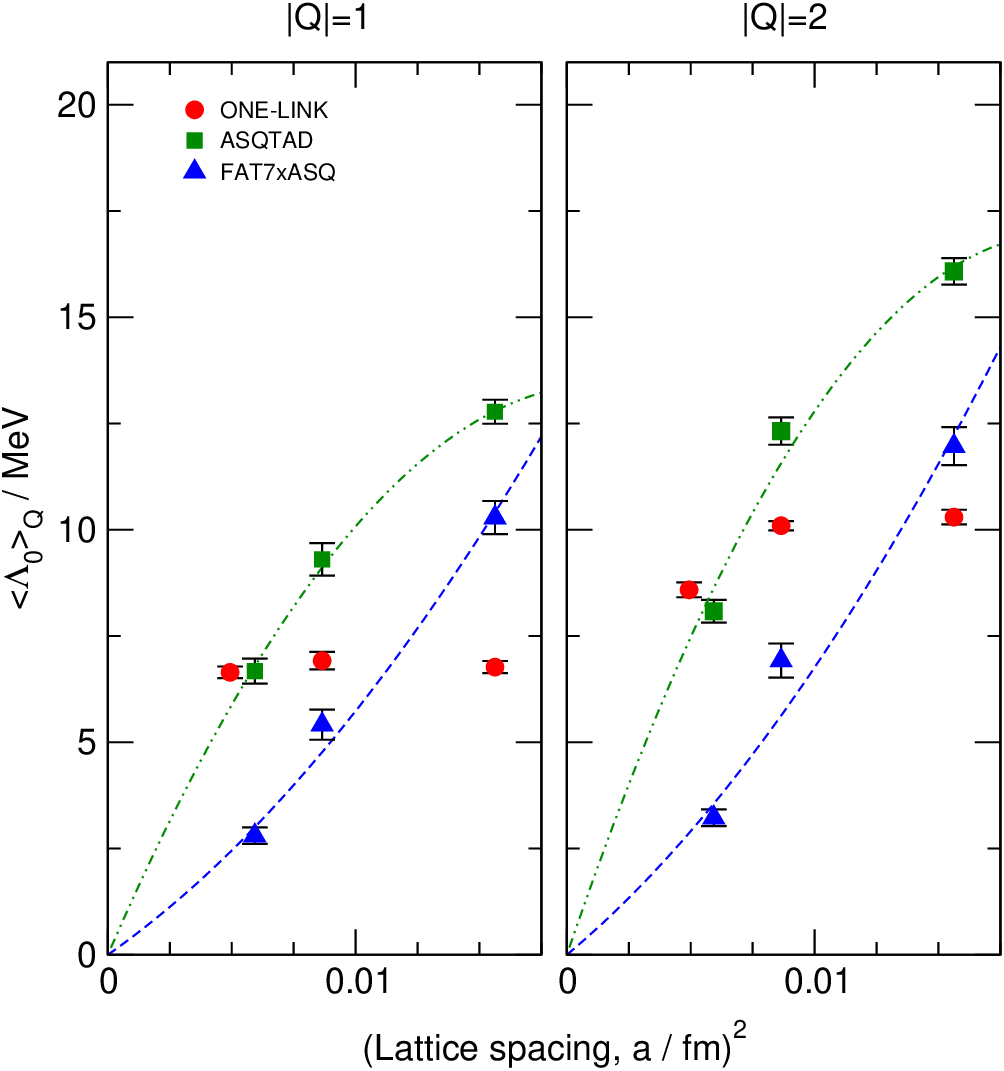,height=6.5cm}
\epsfig{file=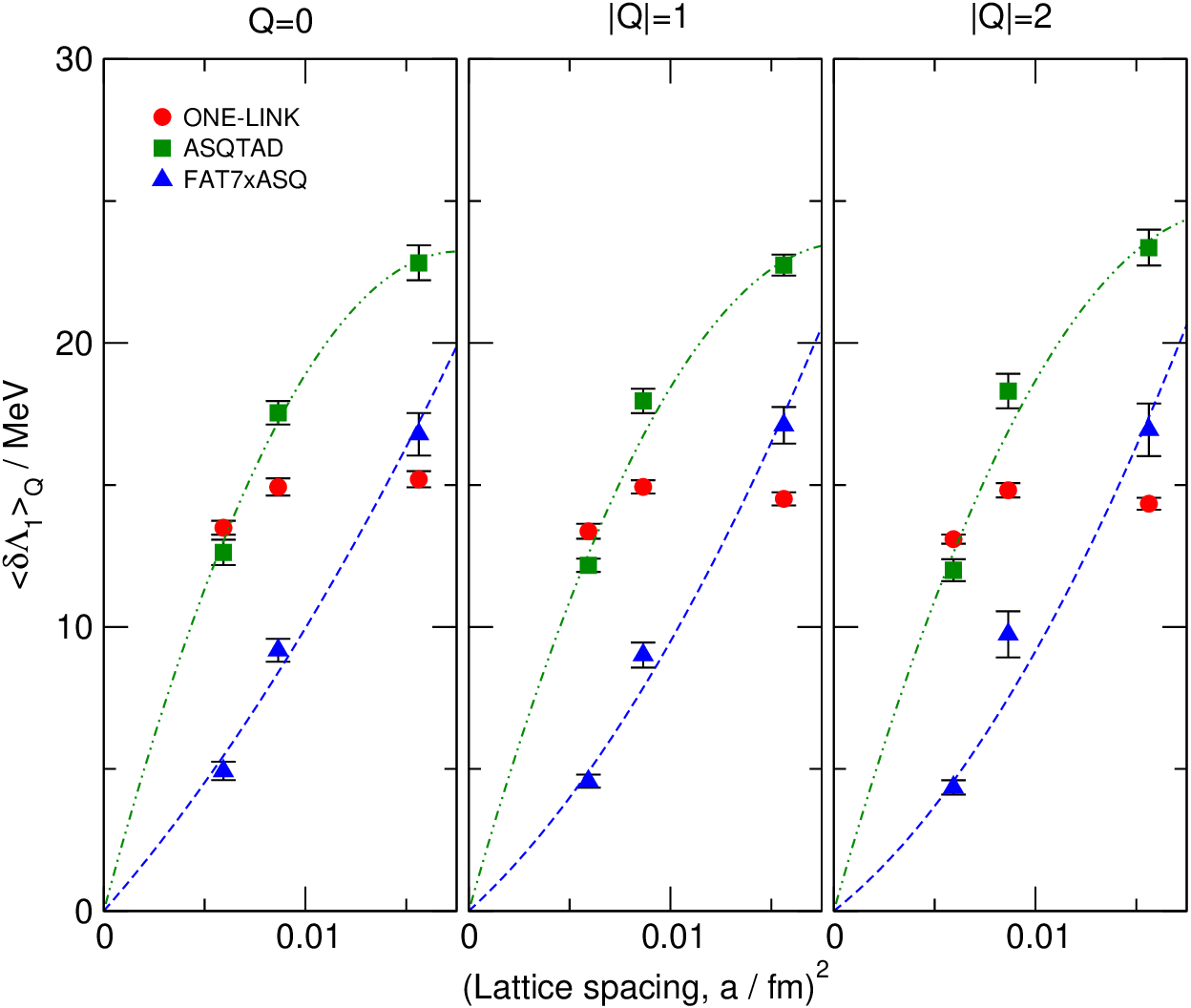,height=6.5cm}
\end{flushright}
\vspace{-8mm}
\caption{Same as Fig.\,8, but separated according to $|q|$ and with much better
statistics. Figure from \cite{Follana:2005km}.}
\label{fig_follana2}
\end{figure}

The last point to be covered is the evidence that the staggered taste symmetry
violation might indeed be an $O(a^2)$ cut-off effect.
A first attempt was made in \cite{Durr:2004as} where the taste splitting is
defined through the ``spread'' of a staggered quartet that is to mimic a
(single) overlap mode.
Plotting the splitting versus $a^2$, everything seems consistent with the
hypothesis that it might disappear in proportion to $a^2$ [with filtering,
without one is certainly not in the scaling regime], see
Fig.\,\ref{fig_duhowe2}.
In \cite{Follana:2005km} these findings have been corroborated with much better
statistics (see Fig.\,\ref{fig_follana2}) and in \cite{Talk_Maresca} a more
physical definition of the taste splitting has been found to lead to the same
overall picture.

In summary, investigations in 4D have so far revealed evidence in favor of
the ``rooting trick''.
The problem is just that even the most convincing numerical evidence does not
amount to a proof.

%%%%%%%%%%%%%%%%%%%%%%%%%%%%%%%%%%%%%%%%%%%%%%%%%%%%%%%%%%%%%%%%%%%%%%%%%%%%%%%

\section{Interacting theory: $\ch_\mr{sca}\,,\ch_\mr{top}$ in 2D}

The main reason for working in 2D is that for a large number of observables
a rather impressive statistical precision can be reached.
Hence, if there is a problem with the staggered approach, one might think that
it is most likely to be detected in such a setup.

In 2D a square root of the determinant is used for an odd number of dynamical
fermions, since the staggered action yields 2 fermions in the continuum.
Results below are in the (generalized) Schwinger model \cite{Schwinger:1962tp},
i.e.\ QED in 2D with $\Nf$ massive fermions.
Note that the fundamental charge $g$ has the dimension of a mass and does
not run.
Therefore, $g$ may be used to set the scale, and in this case the relation
$\be\!=\!1/(ag)^2$ holds exact.
For $\Nf\!=\!1$ there is the analytic result \cite{Schwinger:1962tp}
\beq
\lim_{m\downarrow0}{\<\psb\ps\> \ovr g}={e^{\ga} \ovr 2\pi^{3/2}}=0.1599...
\label{res_schwinger}
\eeq
which is due to the axial anomaly and does not signal spontaneous symmetry
breaking.

The challenge is to see whether one may directly reproduce
(\ref{res_schwinger}) with staggered fermions.
To this aim we sample quenched, compute all eigenvalues (both of $\Dst$ and
$\Dov$ to have a cross check, and in either case with several filterings) and
determine the observable in the analysis program via
\bea
%{\ch_\mr{sca}^\mr{ov}\ovr g}=-{\sqrt{\be}\ovr V}\<\psb\ps\>
%&\;,\qquad&
%{\ch_\mr{sca}^\mr{st}\ovr g}=-{\sqrt{\be}\ovr2V}\<\chb\ch\>
%\\
{\ch_\mr{sca}^\mr{ov}\ovr g}={\sqrt{\be}\ovr V}\,
{\<\det^{\Nf}(D_m^\mr{ov})\sum{1\ovr\hat\la+m}\>\ovr
\<\det^{\Nf}(D_m^\mr{ov})\>}
&\;,\qquad&
{\ch_\mr{sca}^\mr{st}\ovr g}={\sqrt{\be}\ovr2V}\,
{\<\det^{\Nf/2}(D_m^\mr{st})\sum{1\ovr\la+m}\>\ovr
\<\det^{\Nf/2}(D_m^\mr{st})\>}
\label{def_cond1}
\\
\det(D_m^\mr{ov})=\prod((1\!-\!{m\ovr2})\la+m)
&\;,\qquad&
\det(D_m^\mr{st})=\prod(\la+m)
\label{def_cond2}
\eea
where $\hat\la\!=\!(1/\la-1/2)^{-1}$ denotes the stereographically mapped
overlap eigenvalues.
The prefactor ${1\ovr2}$ in the staggered expression is designed to compensate
for the 2 tastes contributing; this is the valence counterpart to the
``rooting trick'' in the sea sector.
The bare condensates (\ref{def_cond1}) have a logarithmic divergence of type
$m/g\cd\log(ma)$, i.e.\ the free case needs to be subtracted, but there is no
multiplicative renormalization.
For technical details see \cite{Durr:2003xs,Durr:2004ta}.
All plots below will have $m_\mr{sea}\!=\!m_\mr{val}$, i.e.\ at least with
$D\!=\!\Dov$ the theory is exactly unitary (at finite lattice spacing).

\begin{figure}
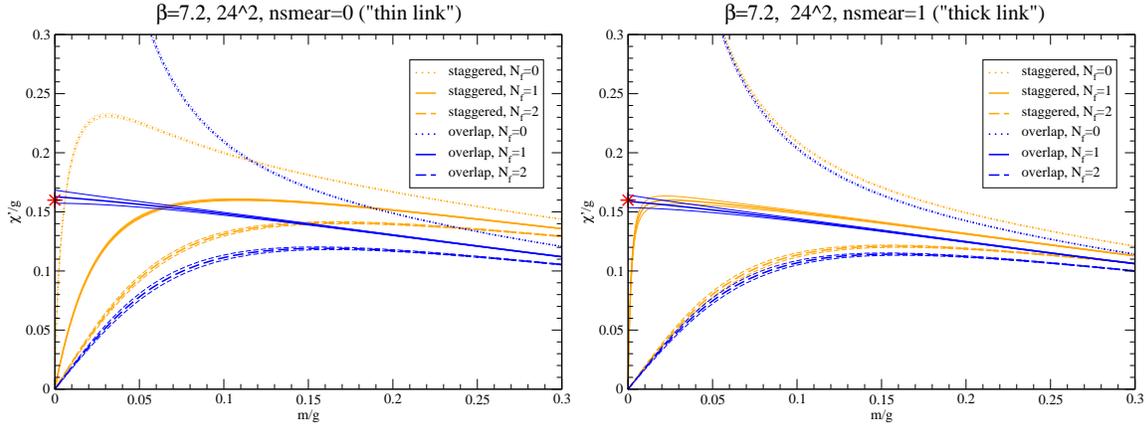

\epsfig{file=stagissues.figs/smrw_anacond0.eps,width=7.5cm}
\epsfig{file=stagissues.figs/smrw_anacond1.eps,width=7.5cm}
\vspace{-8mm}
\caption{Scalar condensate versus quark mass (everything in physical units)
in the Schwinger model, with $\Dst$ and $\Dov$, with $\Nf\!=\!0,1,2$,
without (left) or with filtering (right). Figure (with modification) from
\cite{Durr:2003xs}.}
\label{fig_cond}
\end{figure}

Given this setup it is clear that we can draw ``continuous'' $1\si$
(statistical) error bands for the subtracted condensate $\ch_\mr{sca}'/g$ as
a function of $m/g$.
The volume is chosen such that the Compton wavelength of the lightest degree
of freedom in the chiral limit for $\Nf\!=\!1$ fits $4-5$ times into the box.
The first observation in Fig.\,\ref{fig_cond} is that our overlap data
satisfy the continuum expectation
\beq
\ch_\mr{sca}'/g\propto
\left\{
\begin{array}{ll}
g/m&(\Nf\!=\!0)\\
\mr{const}&(\Nf\!=\!1)\\
m/g&(\Nf\!=\!2)
\end{array}
\right.
\label{expectation}
\eeq
at least on a qualitative level.
Moreover, the overlap results with and without filtering look almost
identical.
And finally, in the chiral limit the 1-flavor result is consistent with
the exact value (\ref{res_schwinger}).
With staggered fermions the picture looks rather different.
The most obvious problem is that in the chiral limit these condensates tend to
zero for all $\Nf$, in marked contrast to (\ref{expectation}).
However, at the coupling used a single filtering step suffices to dramatically
improve the situation -- for not-too-small quark masses at least.
Still one has $\;\lim_{m\to0}\ch_\mr{sca}^{\prime\;\mr{st}}/g\!=\!0\;$
at finite lattice spacing (even for $\Nf\!=\!0$ where the slope is so steep
that already the first pixel is off scale), but the point where this happens
has been shifted to the left.
This is in line with our previous observation that the splitting within
staggered eigenvalue doublets gets dramatically reduced through filtering
(cf.\ Fig.\,\ref{fig_spec_sm} and Ref.\,\cite{Durr:2003xs}).
However, there is a large numerical factor relating the intra-taste
eigenvalue splitting and the breakdown mass, and this ought to be understood
before one claims a relationship.
The main question that emerges from the r.h.s.\ of Fig.\,\ref{fig_cond} is
whether one can reproduce (\ref{res_schwinger}) with staggered fermions if one
\emph{first\/} takes the continuum limit (at finite $m/g$), pushing to the
chiral limit in a \emph{second\/} step.
It turns out that this is indeed the case, viz.
\bea
\lim_{a\to0}\lim_{m\to0}
 {\ch_\mr{sca}^{\prime\;\mr{ov}}(m/g,a^2)\ovr g}={e^{\ga}\ovr2\pi^{3/2}}
&\;,\qquad&
\lim_{a\to0}\lim_{m\to0}
 {\ch_\mr{sca}^{\prime\;\mr{st}}(m/g,a^2)\ovr g}=0
\label{non_comm1}
\\
\lim_{m\to0}\lim_{a\to0}
 {\ch_\mr{sca}^{\prime\;\mr{ov}}(m/g,a^2)\ovr g}={e^{\ga}\ovr2\pi^{3/2}}
&\;,\qquad&
\lim_{m\to0}\lim_{a\to0}
 {\ch_\mr{sca}^{\prime\;\mr{st}}(m/g,a^2)\ovr g}={e^{\ga}\ovr2\pi^{3/2}}
\label{non_comm2}
\eea
where every identity except the one with the exact zero is to be read as
``our data are consistent with the hypothesis that ... '' \cite{Durr:2004ta}.
Thus there is a \emph{non-commutativity\/} phenomenon specific to the
staggered action.
In other words, for this theory the staggered action is not in the right
universality class, if one implements $\ch_\mr{sca}/g$ through
(\ref{def_cond1}) and insists on a particular ``bad'' order of limits.
Perhaps more surprising to staggered skeptics is that the ``good'' order of
limits (first $a\!\to\!0$, then $m\!\to\!0$) brings it to the right
universality class.
This means that staggered quarks --~with the help of an extra guiding
principle~-- do perceive the \emph{chiral anomaly\/} \cite{Durr:2004ta}.

\begin{figure}
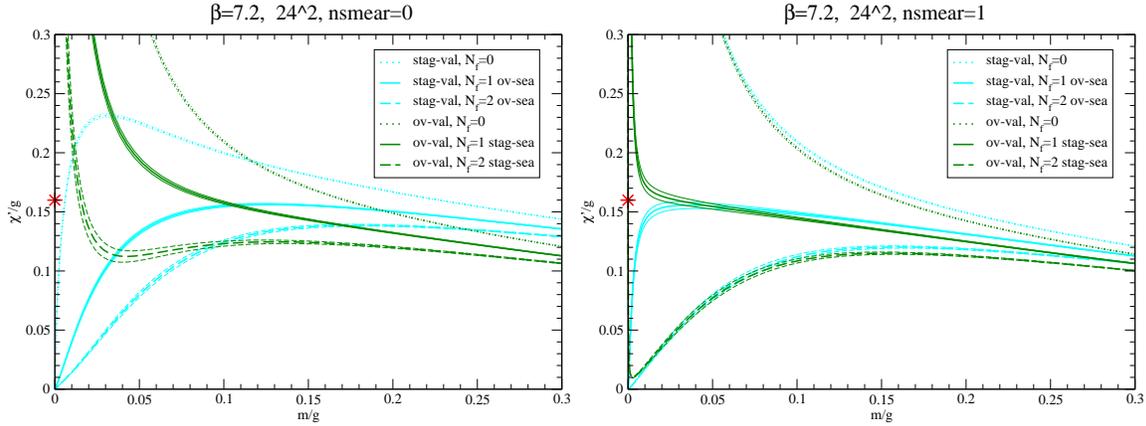

\epsfig{file=stagissues.figs/smrw_anahybr0.eps,width=7.5cm}
\epsfig{file=stagissues.figs/smrw_anahybr1.eps,width=7.5cm}
\vspace{-8mm}
\caption{Same as Fig.\,10, but for the formulations
\emph{overlap valence on staggered sea\/} (dark) and
\emph{staggered valence on overlap sea\/} (light), without (left) and with
filtering (right). Figure (with modification) from \cite{Durr:2003xs}.}
\label{fig_hybr}
\end{figure}

What makes staggered quarks fail, in this specific case, with the ``bad''
order of limits ?
We did two experiments.
Already in \cite{Durr:2003xs} we asked what happens if one combines staggered
sea with overlap valence quarks or vice versa (BTW, \cite{Durr:2003xs}
contains the first published ``hybrid action'' test).
In such a setup one has the ``taste reduction'' problem only in one of the
two sectors.
We restrict ourselves to $m_\mr{sea}\!=\!m_\mr{val}$.
As shown in Fig.\,\ref{fig_hybr} the results with overlap valence quarks on
staggered sea look similar to the genuine staggered data shown before.
Again, the curves tend to zero below some ``critical'' quark mass, and the
latter happens to be more pronounced (and smaller) with filtering than without.
An interesting observation concerns the reverse combination, staggered valence
quarks on overlap sea (which in 4D would be a crazy thing to do).
Here, the curves blow up at the same ``critical'' mass.
At least in this example the notion of exact chiral symmetry in either the
valence or the sea sector alone does not enable one to go closer to the chiral
limit (at fixed $a$) than with the generic ``staggered on staggered''
combination.
This is potentially bad news for modern attempts to evaluate
overlap/domain-wall valence quarks on ``cheap'' staggered ensembles
(see \cite{Bowler:2004hs,Golterman:2004mf} for a discussion of specific issues
and \cite{Bar:2002nr,Bar:2005tu} for the effective theory).
For our problem the bottom line is that the failure of the staggered action in
the 1-flavor theory with the ``bad'' order of limits cannot be attributed to
the sea or valence sector alone.

\begin{figure}
\hfill
\epsfig{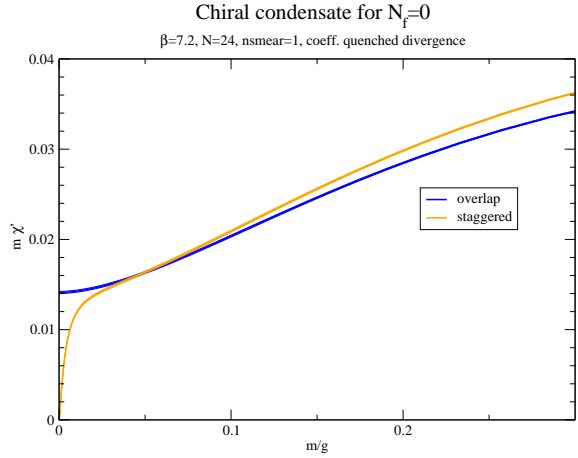}
\hfill
\vspace{-3mm}
\caption{Attempt to ``zoom in'' on the coefficient of the divergence in the
$\Nf\!=\!0$ case. Overlap fermions yield the right result, regardless whether
$m\!\to\!0$ or $a\!\to\!0$ is taken first. With staggered fermions another
non-commutativity shows up which does not involve a rooted determinant.
Figure previously unpublished.}
\label{fig_quen}
\end{figure}

The second experiment is the result of a discussion with Claude Bernard.
To find out whether the non-commutativity (\ref{non_comm1},\ref{non_comm2})
is causally connected to the ``rooting trick'' in the 1-flavor theory with
staggered fermions, we checked whether one can compute the coefficient of
the quenched divergence in (\ref{expectation}) with staggered fermions.
In Fig.\,\ref{fig_quen} the product $m\ch_\mr{sca}^{\prime\;\mr{st/ov}}$ is
plotted versus $m/g$.
With overlap quarks the order of the chiral and the continuum limit is
irrelevant.
With staggered quarks taking $m\!\to\!0$ first yields a vanishing result,
hence taking $a\!\to\!0$ in the second step leads to an exact zero.
However, the result is different if one reverses the order of limits.
With $a\!\to\!0$ the ``dip'' in Fig.\,\ref{fig_quen} gets filled in for every
finite $m/g$.
Therefore, taking $m\!\to\!0$ in the second step one ends up with a finite
result (which is conjectured to be identical to the coefficient obtained from
overlap fermions in the combined chiral and continuum limit).

At the moment it is not clear what the precise implication of any of these
experiments is for the case of interest (QCD with $\Nf\!=\!2\!+\!1$ quarks
at their physical masses).
Evidently, the chiral limit at finite lattice spacing seems to be troublesome
for staggered fermions.
This has already been seen by Smit and Vink in Ref.\,\cite{Smit:1986fn} where
they determine (mass dependent) renormalization factors for the topological
charge as seen by the massive $\Dstm$.
These factors tend to diverge for $m\!\to\!0$.
Our contribution was just to promote this into troubles with a real physical
observable.
In any case the bottom line is that the non-commutativity 
(\ref{non_comm1},\ref{non_comm2}) is not tied to the ``rooting trick'' per se
but rather reflects a general mismatch between the sea and valence sector.
This fits the view by Bernard who observes that similar non-commutativities
show up in the effective theory for staggered fermions in 4D with partial
quenching, i.e.\ if the sea and valence quarks have different masses
\cite{Bernard:2004ab}.

\begin{figure}
\epsfig{file=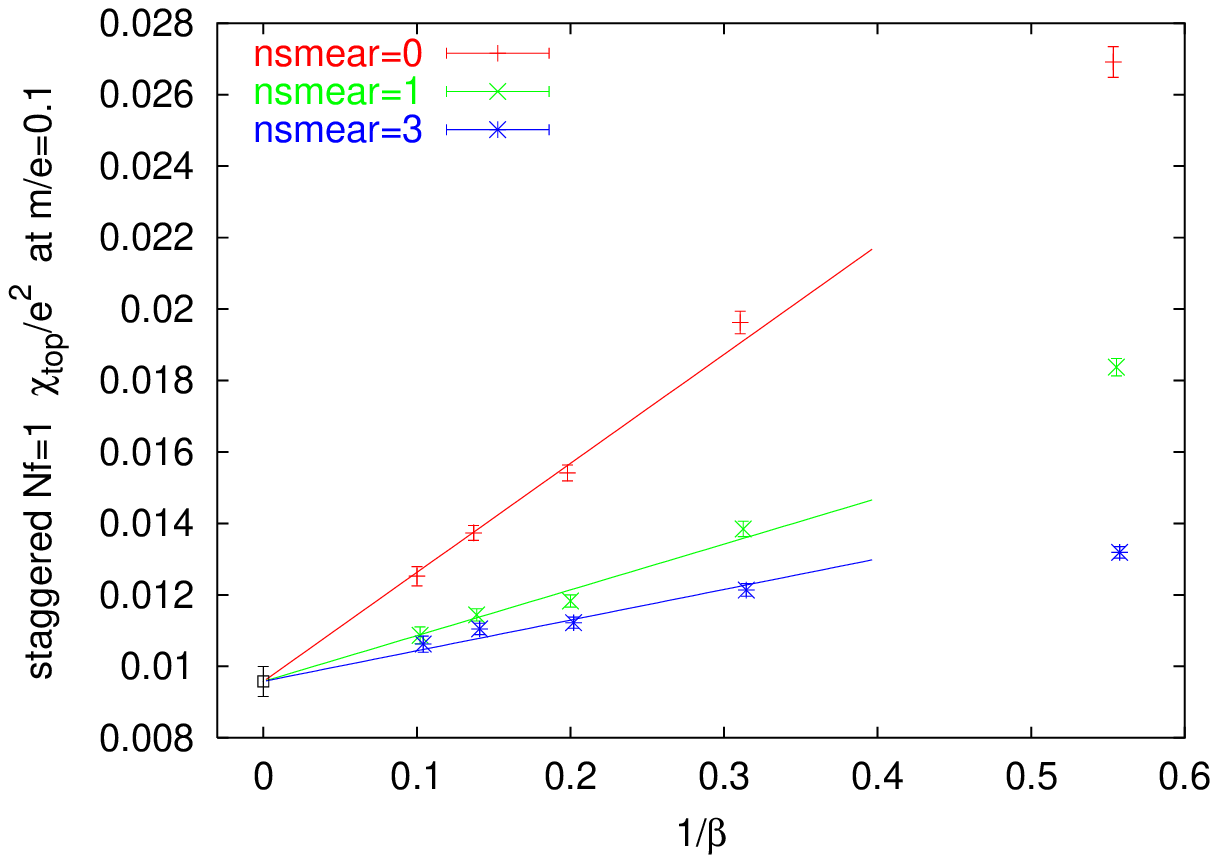,width=7.5cm}
\epsfig{file=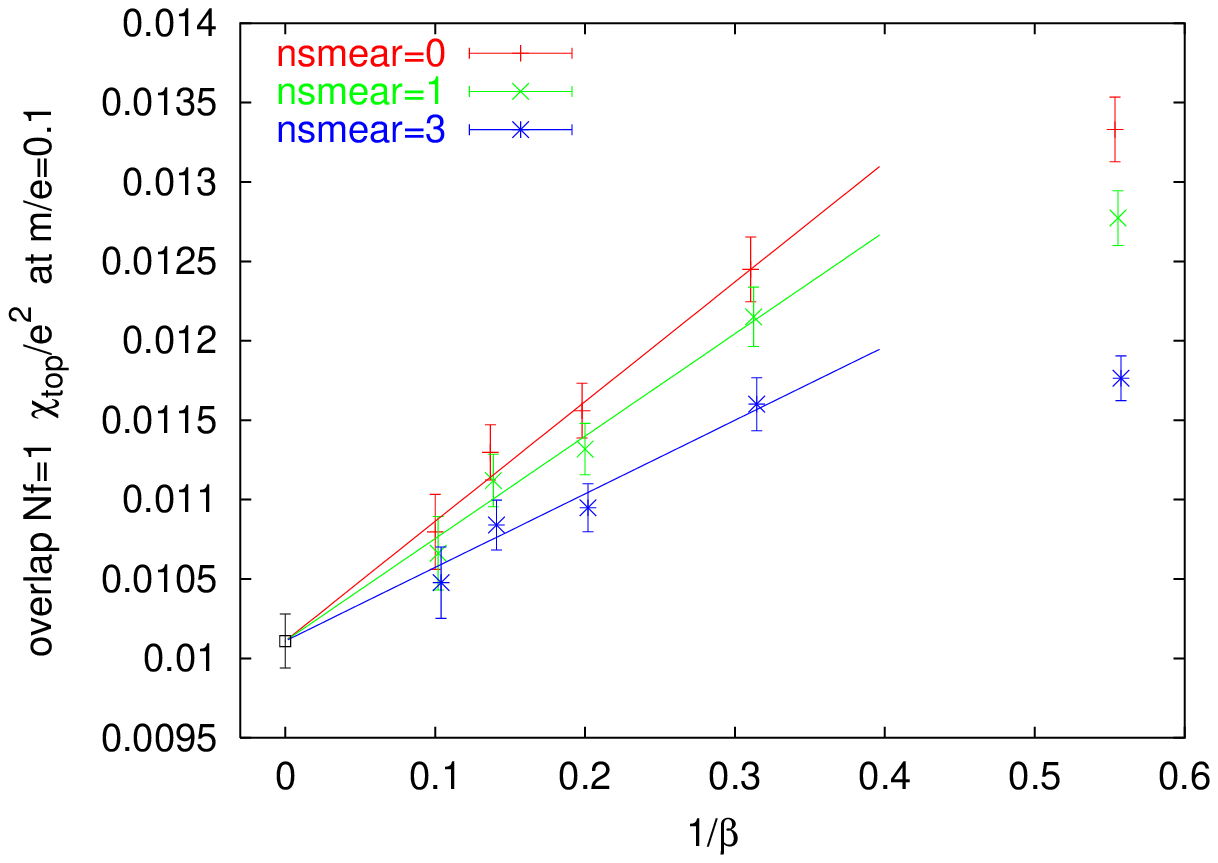,width=7.5cm}
\vspace{-8mm}
\caption{Topological susceptibility $\ch_\mr{top}/g^2$ in the 1-flavor theory
with $\det^{1/2}(\Dstm)$ (left) or $\det(\Dovm)$ (right) at fixed
$m/g\!=\!0.1$, plotted versus $(ag)^2$. The continuum limit seems to
be universal. Figure from\,\cite{Durr:2004ta}.}
\label{fig_susc}
\end{figure}

Finally, I should add that the non-commutativity
(\ref{non_comm1},\ref{non_comm2}) is more the exception than the rule.
We checked a variety of other observables in the Schwinger model and did not
find anything similar.
To give an idea Fig.\,\ref{fig_susc} shows our data for the $\Nf\!=\!1$
topological susceptibility $\ch_\mr{top}/g^2$ at fixed physical quark mass.
Having data with three filtering levels and with several couplings in the
Symanzik scaling regime allows for a fairly safe extrapolation to the
continuum.
The results with the two discretizations seem consistent, and also the exact
continuum result $\lim_{m\to0}\ch_\mr{top}/g^2\!=\!0$ is reproduced with
staggered fermions, regardless whether $m\!\to\!0$ or $a\!\to\!0$ is
taken first \cite{Durr:2004ta}.

%%%%%%%%%%%%%%%%%%%%%%%%%%%%%%%%%%%%%%%%%%%%%%%%%%%%%%%%%%%%%%%%%%%%%%%%%%%%%%%

\section{Correlation of $\det^{1/\Nt}(D_{\mr{st},m})$ and $\det(D_{\mr{ov},m})$}

We have seen, both in 2D and in 4D, that $\Nt$-fold nearly degenerate staggered
eigenvalues manage to ``mimic'' a (single) overlap eigenvalue, e.g.\
$\la_k^\mr{ov}\simeq\ga_k^\mr{st}$ with
$\ga_k^\mr{st}=\ri(-\la_{2k-1}^\mr{st}\la_{2k}^\mr{st})^{1/2}$ in 2D.
Obviously, such a relation cannot extend to the entire spectrum (they are
rather different in the UV, and the number of modes does not match), but the
relevant question is whether it would extend to determinant ratios (from a
configuration $U$ to $U'$) as seen by these formulations, i.e.\ whether
\beq
{\la_1\,\la_2\;...\ovr\la_1'\,\la_2'\;...}\,\Big|_\mr{ov}\;\simeq\;
{\ga_1\,\ga_2\;...\ovr\ga_1'\,\ga_2'\;...}\,\Big|_\mr{st}\;.
\eeq
If such a relation can be shown, specifically if one can prove that the
deviation from an exact identity is caused by $O(a^2)$ cut-off effects,
one would know that the rooted staggered determinant creates the ``same''
ensemble as the overlap action.

\begin{figure}
\epsfig{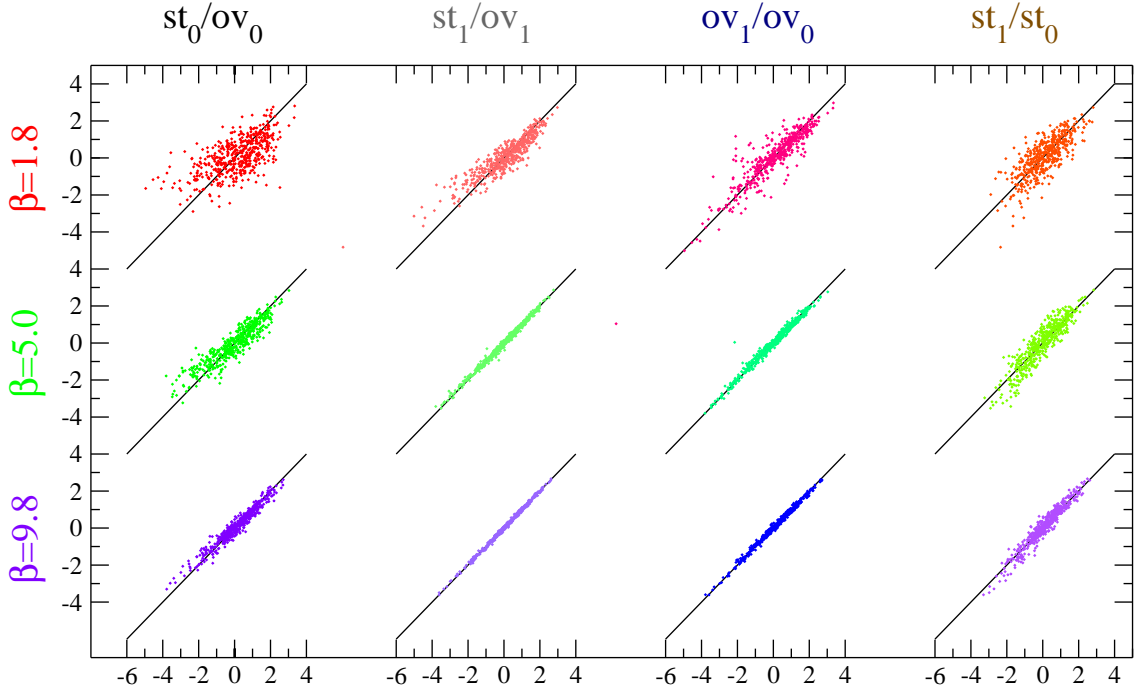}
\caption{Scatter plot of ${1\ovr2}\log\det(D_{\mr{st},m})$ versus 
$\log\det(D_{\mr{ov},m})$ in a box of fixed physical volume and with
$m/g\!=\!0.1$ fixed. The first column is without filtering, the second with.
For comparison the correlation of the overlap determinant with and without
filtering is shown and ditto for the rooted staggered determinant,
where the deviations are known to be $O(a^2)$ effects.
The black lines represent the identity. Figure from \cite{Durr:2004ta}.}
\label{fig_corr}
\end{figure}

In 2D it is rather easy to compute exact determinants, in particular if one has
all eigenvalues on several ensembles in a matched physical volume on disk.
This being the case with our Schwinger model studies, we included a correlation
plot in \cite{Durr:2004ab,Durr:2004ta}, see Fig.\,\ref{fig_corr}.
The sign-flipped contribution to the effective action on typical quenched 
backgrounds, as seen by (rooted) staggered or (single) overlap fermions is
plotted against each other.
In either case the determinant has been  normalized by an artificial reference
configuration which realizes the ensemble mean in both directions, and the
physical box volume and quark mass are the same in all graphs.
On coarse lattices, there is a rather loose correlation, i.e.\ we see a
``cigar'' shaped cloud.
The interesting observation is that the thickness of the ``cigar'' shrinks
on finer lattices, i.e.\ the correlation improves towards the continuum.
This is particularly obvious if the two operators are filtered (second column).
The full line always represents the identity, not a fit.
To interpret this plot it is useful to recall that a correlation  between
$S_G$ and $-\log\det(\Dovm)$ or $-{1\ovr2}\log\det(\Dstm)$ looks rather
similar on coarse lattices, but here the correlation does \emph{not\/} improve
towards the continuum
\footnote{It simply means that there is a genuine unquenching effect that
cannot be mimicked through a shift in $\be$.}.
For comparison the third and fourth columns correlate different filtering
levels for the overlap or staggered action, respectively.
The point is that in these cases the difference is known to be an $O(a^2)$
effect.
It is amusing to see that a filtered overlap determinant creates an ensemble
that is closer to the one from a rooted staggered determinant at the same
filtering level than to the one from an unfiltered overlap determinant.
In summary, the similarity between the second and third column makes it
plausible that $\det^{1/\Nt}(\Dstm)$ and $\det(\Dovm)$ create ensembles that
differ by cut-off effects only (a more quantitative analysis is included
in \cite{Durr:2004ta}).
To my knowledge Fig.\,\ref{fig_corr} contains the strongest evidence
available right now that the overlap operator $\Dovm$ might actually be
(one version of) the desired ``candidate'' action $\Dcam$,
i.e.\ that the conjectured identity
\beq
\det(\Dovm)=\const\cdot\det\nolimits^{1/\Nt}(\Dstm)\,(1\!+\!O(a^2))
\label{conjec}
\eeq
might hold true with $\Nt\!=\!2^{d/2}$ (it is understood that
the two operators have the same renormalized mass).
On the other hand, it could be that all this is just misleading.
In that case the only lesson to be learned would be that numerical evidence
--~no matter how convincing it looks~-- is never good enough to decide on
conceptual issues.

%%%%%%%%%%%%%%%%%%%%%%%%%%%%%%%%%%%%%%%%%%%%%%%%%%%%%%%%%%%%%%%%%%%%%%%%%%%%%%%

\section{Staggered Chiral Perturbation Theory}

The last piece of evidence in favor of the rooting procedure comes from fitting
hadron data to functional forms specific to the the low-energy effective theory
for staggered quarks.
For a general review on effective theories at finite lattice spacing
I refer to \cite{Bar:2004xp}.

Let me first recall some of the foundations of standard Chiral Perturbation
Theory (XPT) as formulated by Gasser and Leutwyler
\cite{Gasser:1983yg,Gasser:1984gg}.
The key observation is that in QCD at low energy (say below $1\GeV$)
chiral symmetry seriously constrains the interaction of ``pions'' ($\pi,K,\et$)
with each other and with heavier degrees of freedom (e.g.\ nuclei, see the
talk by U.-G.\ Meissner \cite{Talk_Meissner}).
The way how this is implemented is through the Goldstone theorem
\cite{Goldstone:1961eq,Goldstone:1962es}.
More specifically, they start from the fundamental assumption that in QCD with
$\Nf$ light quarks (the focus is on the meson sector with $\Nf\!=\!2,3$)
the pattern of spontaneous symmetry breaking (SSB) is
\beq
SU(\Nf)_L \times SU_R(\Nf)_R \;\longrightarrow\; SU(\Nf)_V
\label{ass_xpt}
\eeq
and build the most general effective Lagrangian consistent with this (and
a few more constraints like Lorentz covariance and parity conservation).
In this framework an amplitude like, e.g.\ the $I\!=\!0$ $\pi\pi$ forward
scattering amplitude [$s\!=\!0$, $\nu\!=\!(t\!-\!u)/(4\Mpi)$] has a
low-momentum expansion
\beq
F_{\pi\pi}(\nu)=
T^{I=0}(0,2\Mpi(\Mpi\!+\!\nu),2\Mpi(\Mpi\!-\!\nu))=
-{\Mpi^2\ovr\Fpi^2}\,+\,O(p^4)
\label{xpt_amp}
\eeq
in powers of $p^2$.
As announced above, (\ref{xpt_amp}) tends to zero for $\Mpi^2,\nu\to0$.
The counting rule $p^2\!\sim\!m$ (where $m$ is identified with $\Mpi^2$) is
dictated by the reaction to infinitesimally small breaking terms.

\begin{figure}
\epsfig{file=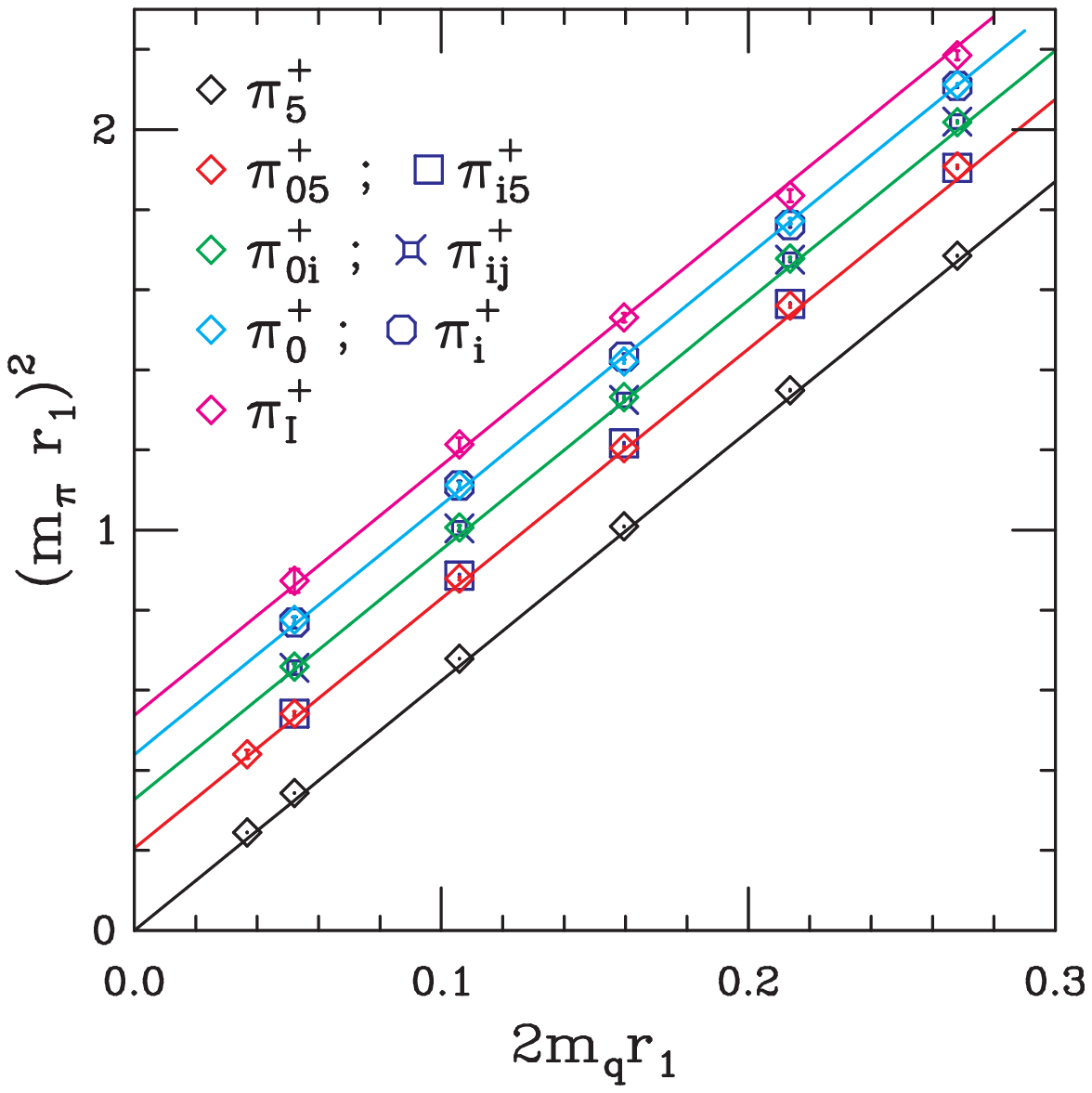,
width=7cm,angle=-90}
\hspace*{2mm}
\epsfig{file=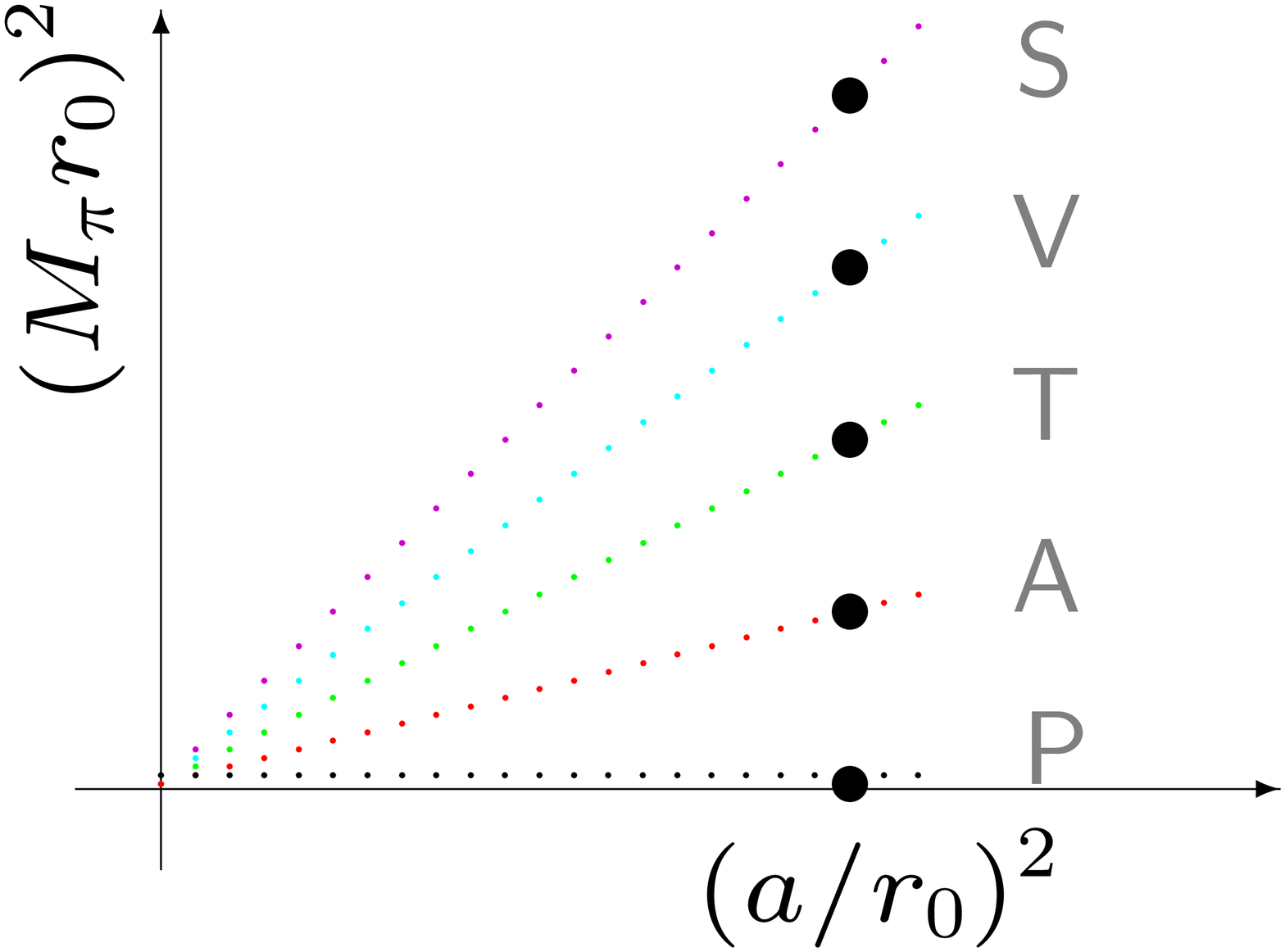,width=5.7cm,angle=-90}
\caption{Left: $\Mpi^2$ versus $m$ for Goldstone pions (lowest line)
and ``non-Goldstone'' pions (in parallel). Figure from
\cite{Aubin:2004fs}. Right: $\Mpi^2$ versus $a^2$ for Goldstone pions (``P'')
and ``non-Goldstone'' pions (``A,T,V,S'').}
\label{fig_leesharpe}
\end{figure}

Given this background, it is clear that an extension of XPT to staggered
quarks must build on an analysis how the broken taste symmetry at finite
lattice spacing affects the pion mass.
This was done in a paper by Lee and Sharpe \cite{Lee:1999zx}, and
Fig.\,\ref{fig_leesharpe} illustrates their main findings.
The generalized ``pions'' built, say, from an up-quark and a down-antiquark
fall into different taste irreps $\bar d(\gaf\!\otimes\!\ta)u$.
Only the taste pseudoscalar ($\ta\!=\!\ta_5$, ``P'' in
Fig.\,\ref{fig_leesharpe}) is massless in the chiral limit at finite lattice
spacing; this state is to be identified with the physical $\pi^+$.
All taste non-P states (``A,T,V,S'' in Fig.\,\ref{fig_leesharpe}) stay
massive in the chiral limit; the excess of their squared masses over the
$\ta\!=\!\ta_5$ partner is nearly independent of the quark mass.
Physical correlators project on taste pseudoscalars in the asymptotic states,
but in intermediate loops all taste irreps do, of course, contribute.
Even with intermediate flips into unphysical tastes being unavoidable, the
continuum limit might be OK, if the contribution from physical intermediate
states is far less than what continuum XPT would predict.
The non-P mass excess diminishes like $a^2$ (see Fig.\,\ref{fig_leesharpe}).
Hence a scheme in which a cut in a 4-point function is built up from several
contributions, each of which has a fraction of the continuum weight, would be
OK if the fractions add up to 1.
In fact, such a desired behavior is built into the effective theory via an
enlarged symmetry group [and a dedicated prescription].
The key assumption is as follows.
With $\Nf$ flavors of ($4$-taste) quark fields the pattern of SSB is
\beq
SU(4\Nf)_L\!\times\!SU(4\Nf)_R\,\to\,SU(4\Nf)_V
\label{ass_sxpt}
\eeq
leading to $16\Nf^2\!-\!1$ pseudo-Goldstone bosons, collected in the
$12\!\times\!12$ matrix (for $\Nf\!=\!3$)
\beq
U=e^{\ri\Phi/f}
\qquad\mbox{where}\qquad\qquad
\Phi=\left(\!\!
\begin{array}{ccc}
\Phi_u\!&\!\pi^+\!&\!K^+\\
\pi^-\!&\!\Phi_d\!&\!K^0\\
K^-\!&\!\bar K^0\!&\!\Phi_s
\end{array}
\!\!\right)=
\sum_{a,b=1}^{9,16}\Phi^{ab}{\la^{\!a}\ovr2}T^b
\label{sxpt_matrix}
\eeq
that transforms as $\,U\to V_L^{}UV_R\dag\,$ under chiral rotations with
unitary $V_L,\!V_R$.

Staggered Chiral Perturbation Theory (SXPT) is the systematic extension of
the Lee-Sharpe analysis to higher loop order, worked out in detail by
Aubin and Bernard \cite{Aubin:2003mg,Aubin:2003uc}.
The counting rule $p^2\!\sim\!m\!\sim\!a^2$ follows from the discussion
above and the leading order Lagrangian reads
\beq
L={f^2\ovr8}\trc(\pa_\mu U \pa_\mu U\dag)-{\Sigma\ovr2}\trc(MU\!\!+\!\!MU\dag)
+{2m_0^2\ovr3}(\Phi_{u,\mr{TS}}\!+\!\Phi_{d,\mr{TS}}\!+\!\Phi_{s,\mr{TS}})
+a^2V_\mr{TB}
\label{sxpt_lagrangian}
\eeq
with $f\!\simeq\!122\MeV$ the pion decay constant in the chiral limit,
$\Sigma\!\simeq\!(270\MeV)^3$ the chiral condensate
[in ($\overline{\mr{MS}},2\GeV$) conventions] and
$M\!=\!\mr{diag}(m_u, ... , m_d, ... , m_s, ... )$ the 
$12\!\times\!12$ quark mass matrix.
From a look at (\ref{sxpt_matrix},\ref{sxpt_lagrangian}) it is clear that some
details of the theory are rather different from ordinary XPT.
The flavor sum in (\ref{sxpt_matrix}) extends up to 9, not just 8 (here,
$\la^9\!=\!\sqrt{2/3}\,I$).
This is necessary, since (\ref{ass_sxpt}) yields $12^2-1=143$ 
Goldstone and non-Goldstone bosons.
With only $8\cd16=128$ terms the dimensionality of the tangent space would
be too small.
On the other hand, with $9\cd16=144$ terms there is space for one extra
degree of freedom.
This extra degree, the physical $\et'$, gets lifted through the piece with
the $m_0^2$ prefactor in (\ref{sxpt_lagrangian}) which is designed to affect
only the combined flavor and taste singlet state.
That the $\et'$ is an explicit degree of freedom creates a similarity with
quenched or partially quenched XPT, even for $m_\mr{sea}\!=\!m_\mr{val}$.
The taste breaking potential $V_\mr{TB}$ in (\ref{sxpt_lagrangian}) implements
the excess mass of the non-Goldstone pions that has been illustrated in
Fig.\,\ref{fig_leesharpe}.
With this setup several important quantities have been calculated at 1-loop
level, e.g.\ for $\Mpi,f_\pi,M_K,f_K$ it is known how they react to changes
in $m$ and $a$ \cite{Aubin:2003mg,Aubin:2003uc}.

The SSB pattern (\ref{ass_sxpt}) has been declared an assumption.
Here, I would like to emphasize that the analogous pattern (\ref{ass_xpt})
remains an assumption in ordinary XPT, too.
Of course, Gasser and Leutwyler mention that there is a wealth of evidence
[from pions in the real world being so light, and even from the lattice]
that the breaking (\ref{ass_xpt}) is actually realized in nature.
The point is that their argument is evidence based, there is no proof that
standard XPT is the correct low-energy theory of QCD.
With SXPT the situation is rather analogous.
One can compare the predictions to lattice data and see whether they fit.
The MILC collaboration has done this in great detail, performing correlated
fits in various circumstances (see e.g.\ \cite{Aubin:2004fs}).
Given the large number of free parameters, it is not so surprising that they
get horrific covariance matrices, but in terms of the $\ch^2/\mr{d.o.f.}$
obtained their fits are quite acceptable.
Perhaps the most significant observation is that they cannot fit the taste
pseudoscalar part of their data with continuum XPT, it is crucial to analyze
their data in the full SXPT framework.
Another point covered in \cite{Aubin:2004fs} is that they generalize their
formulae to allow for an arbitrary number of remaining tastes per flavor
(not just $1$).
Fitting some of their lighter data with this extra parameter, they obtain
$1.28(12)$ as the preferred number.
More detailed tests have been proposed by Sharpe and van de Water
\cite{Sharpe:2004is}, and experiences in more exotic sectors have been
reported at this conference \cite{Talk_Gregory,Talk_Prelovsek}.
At this point there is a huge body of evidence that SXPT allows for an
excellent description of a wealth of lattice data with the rooted staggered
determinant.
In principle there is, of course, the possibility that the ``rooting trick''
is incorrect and SXPT just a stunningly accurate model.
Personally, I do not consider any accidental agreement of two flawed theories
very likely, and this is why I rate the overall outcome of the MILC fits
good evidence in favor of both the ``rooting trick'' and the SSB pattern
(\ref{ass_sxpt}).
Still, the generic comment applies that even the best numerical evidence does
not amount to a mathematical proof.

%%%%%%%%%%%%%%%%%%%%%%%%%%%%%%%%%%%%%%%%%%%%%%%%%%%%%%%%%%%%%%%%%%%%%%%%%%%%%%%

\section{Summary}

The question whether in QCD with $\Nf\!=\!2\!+\!1$ light quarks the staggered
action yields the right continuum limit remains unsolved.
Instead of drawing any conclusion, let me just summarize:
\begin{enumerate}
\itemsep-0mm
\item
Full QCD with $\Nf\!=\!2\!+\!1$ staggered fermions is controversial, since
the Boltzmann weight $\det^{1/4}(D_\mr{st})$ assumes a taste symmetry
which is only approximate.
\item
Formally, the taste symmetry breaking is due to irrelevant interaction terms
and should thus go away in the continuum limit.
\item
Weak coupling, filtering, RG blocking reduce the taste splitting and give
staggered quarks more appealing features, but there is no guarantee that
no trace is left in the continuum.
\item
One legal 1-flavor $\Dcam$ operator with
$\det(\Dcam)=\const\cd\det^{1/4}(\Dstm)\,(1\!+\!O(a^2))$
is sufficient to re-interpret existing staggered ensembles as being generated
with a \emph{local\/} action.
\item
Even if such a $\Dcam$ exists, the problem of (exact) \emph{unitarity\/} in the
fundamental theory remains when sticking to staggered spectroscopy.
If the construction is exact, i.e.\ if there is no $O(a^2)$ term above,
the problem is easily overcome by using this $\Dcam$ in the valence sector.
\end{enumerate}

\bigskip

{\bf Acknowledgments}:
I would like to thank Christian Hoelbling and Urs Wenger for a series of
enjoyable collaborations and Christian for specific help in preparing
the talk and for allowing me to show Fig.\,12.
In addition, I benefitted from discussions with G.\,Bali, C.\,Bernard,
G.\,Colangelo, C.\,DeTar, K.\,Jansen, A.\,Kronfeld, F.\,Maresca,
F.\,Niedermayer, Y.\,Shamir and S.\,Sharpe.

%%%%%%%%%%%%%%%%%%%%%%%%%%%%%%%%%%%%%%%%%%%%%%%%%%%%%%%%%%%%%%%%%%%%%%%%%%%%%%%

\end{document}